\pdfoutput=1
\documentclass[runningheads]{llncs}

\usepackage{eccv}

\usepackage{eccvabbrv}
\usepackage{graphicx}
\usepackage{booktabs}
\usepackage{pifont}
\usepackage{xcolor}

\usepackage{caption} % 必须引入，用于在同一个环境中使用不同类型的标题
\usepackage{booktabs}
\usepackage{graphicx}
\usepackage{subcaption} % 确保在导言区引入
\usepackage{wrapfig}
\usepackage{amsmath}
\usepackage{amssymb}
\graphicspath{{./}}
\usepackage{hyperref}

\usepackage{orcidlink}

\begin{document}

% ---------------------------------------------------------------
% Title and Authors
\title{HoloTetSphere: Unified TetSphere Mesh Reconstruction for Physical Simulations}

\titlerunning{HoloTetSphere}

\newcommand\samethanks[1][\value{footnote}]{\footnotemark[#1]}

\author{YaQiao Dai\inst{1}\thanks{Equal contribution.} \and
Renjiao Yi\inst{1}\samethanks \and
Zhirui Gao\inst{1} \and
Wei Chen\inst{1} \and
Kai Xu\inst{2} \and
Chenyang Zhu\inst{1}\thanks{Corresponding author. \email{zhuchenyang07@nudt.edu.cn}}}

\authorrunning{Y. Dai et al.}

\institute{National University of Defense Technology, Changsha, China \and
Institute of AI for Industries, Chinese Academy of Sciences, Nanjing, China}

\maketitle

% ---------------------------------------------------------------
% Teaser Figure
\begin{center}
    \includegraphics[width=0.97\textwidth]{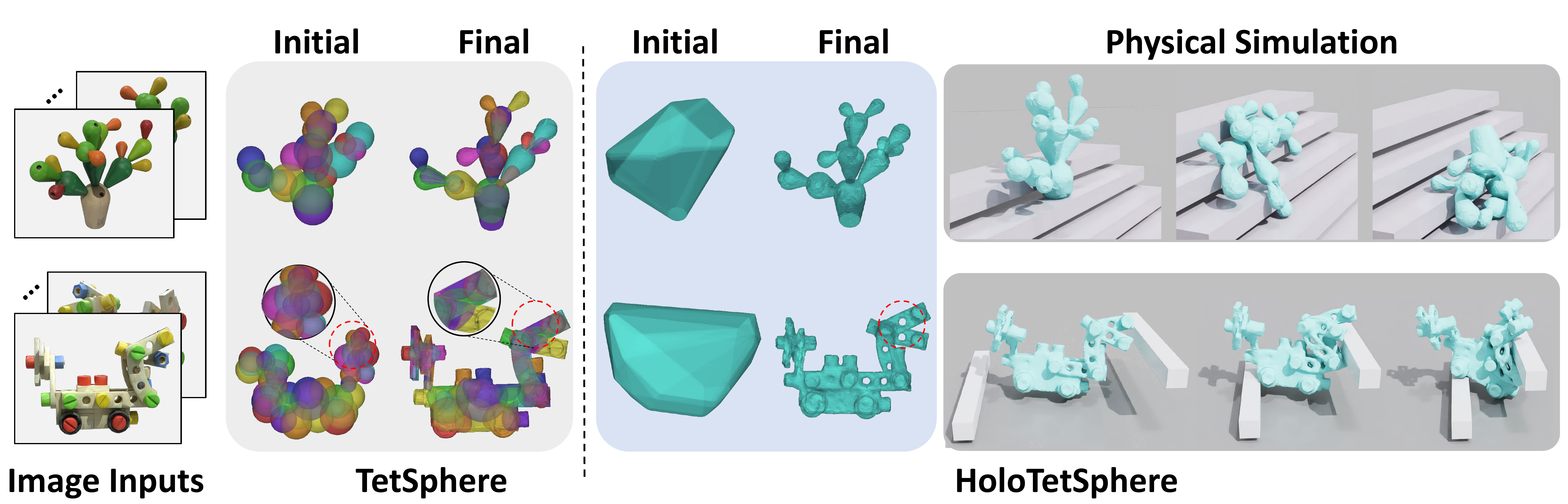}
    \captionof{figure}{Unlike TetSphere's initialization-dependent topology, our method generates holistic tetrahedral meshes through adaptive topology optimization during reconstruction, producing unified and topologically coherent volumetric meshes suitable for downstream physical simulation.}
    \label{fig:teaser}
    \vspace{-10pt}
\end{center}

\begin{abstract}
Standard pipelines for physics-ready 3D reconstruction rely on a decoupled two-stage paradigm: extracting surface geometry followed by an error-prone tetrahedralization process. While recent Lagrangian methods like TetSphere Splatting attempt to bypass this by directly optimizing volumetric primitives, their homeomorphic constraints prevent topology-adaptive optimization. Consequently, they produce disjoint tetrahedra rather than a single connected mesh, rendering the structures unsuitable for further physical simulations. To address this, we propose a topology-adaptive framework for holistic tetrahedral mesh reconstruction through end-to-end topological and geometric optimization. First, by coupling Gaussian spheres to tetrahedral elements and leveraging edge connections, we estimate a continuous opacity field for differentiable element pruning. Next, jointly minimizing mesh smoothing energy and multi-view Gaussian rendering error drives alternating geometric refinement while preserving topological adaptivity. 
Consequently, our approach effectively constructs a unified and topologically coherent tetrahedral mesh. Extensive experiments demonstrate that our method outperforms state-of-the-art techniques by achieving superior geometric accuracy and producing coherent, single-connected tetrahedral meshes, thereby effectively bypassing the error-prone conventional tetrahedralization step for reconstructed surface meshes and streamlining downstream physical simulation.
% First, by coupling Gaussian spheres to tetrahedral elements and leveraging edge connections, we estimate a continuous opacity field to identify and prune unnecessary elements, enabling differentiable topology optimization. Next, by minimizing tetrahedral mesh smoothing energy alongside multi-view Gaussian rendering error, the coupled system performs alternating geometric refinement while preserving topological adaptivity.
\end{abstract}

\section{Introduction}
\label{sec:intro}

Accurate 3D shape modeling serves as a cornerstone for numerous applications ranging from virtual reality to physical simulation and robotics. Current approaches primarily fall into two distinct paradigms: Eulerian and Lagrangian representations. Eulerian methods define geometry on a fixed grid structure, where shape information is encoded at predetermined spatial locations, enabling precise topological control but often at significant computational cost. In contrast, Lagrangian representations track geometry through movable elements that deform with the shape, offering computational efficiency but traditionally suffering from limited reconstruction accuracy due to their discrete and disconnected nature. While Lagrangian approaches excel in computation efficiency, their inability to maintain topological consistency and surface integrity has hindered their adoption in applications requiring physically plausible geometry.

Recent work by Guo \etal~\cite{guo2024tetsphere} introduced TetSphere Splatting, a promising advancement that bridges this gap by incorporating tetrahedral mesh connectivity into the Lagrangian framework. By representing shape as deformable tetrahedral primitives, TetSphere achieves remarkable reconstruction quality while maintaining the computational advantages of Lagrangian methods. This approach successfully generates high-fidelity meshes with inherent connectivity, addressing a critical limitation of previous point-based representations like 3D Gaussian Splatting. However, the integration of tetrahedral meshes within the Lagrangian paradigm introduces a critical constraint: the requirement of homeomorphic mapping during optimization. This constraint prevents true topological adaptation, forcing TetSphere to rely on a fixed set of predefined, disjoint primitives that approximate the target topology through complex preprocessing. Consequently, the method becomes critically dependent on initialization quality, and the resulting disconnected primitives forfeit the physical simulation advantages inherent to unified tetrahedral meshes.

\begin{figure}[t]
  \centering
  \includegraphics[width=0.96\linewidth]{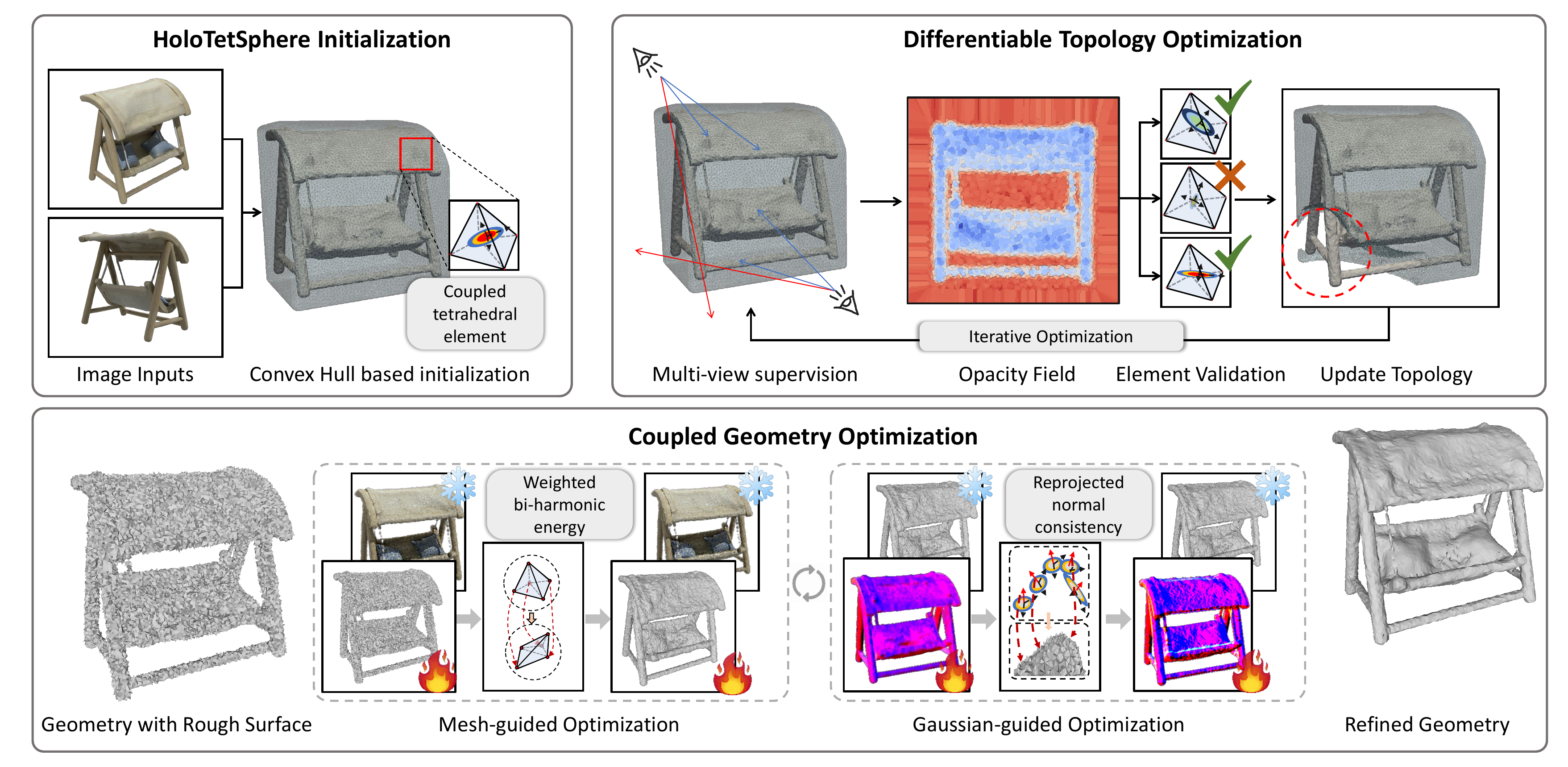}
  % \caption{The framework of HoloTetSphere. It initializes from a coarse, connected tetrahedral mesh covering the object. During optimization, a hybrid representation is adopted, where tetrahedral elements are coupled with Gaussians and optimized in an alternative manner. With a differentiable pruning mechanism, redundant tetrahedral elements can be identified and removed, instead of relying on fixed initializations. Next, we introduce an alternating geometric optimization strategy between tetrahedral elements and their coupled Gaussians. Weighted bi-harmonic energy and re-projected normal consistency are adopted to guide the optimization in these two stages respectively.}
  \caption{Overview of the HoloTetSphere framework. Initialized with a coarse tetrahedral mesh, the system alternately optimizes tetrahedra and coupled Gaussians. A differentiable pruning mechanism dynamically identifies and removes redundant tetrahedra. The alternating optimization stages for tetrahedra and Gaussians are guided by weighted bi-harmonic energy and re-projected normal consistency, respectively.}
  \label{fig:tet_pip}
  \vspace{-10pt}
\end{figure}

To overcome these limitations, we propose a topology-adaptive TetSphere reconstruction framework that achieves holistic mesh generation through end-to-end topological and geometric optimization (\cref{fig:teaser}). As shown in \cref{fig:tet_pip}, our approach begins with a coarse, connected tetrahedral mesh covering the entire target object, then performs adaptive pruning of unnecessary elements guided by multi-view optimization. The key innovation enabling this topological adaptivity is our coupling mechanism: each tetrahedral element is associated with a Gaussian sphere whose position is derived through vertex interpolation of the tetrahedron. 
% Different from optimizing the opacity of each Gaussian sphere independently, we adopt the tetrahedral edges as bridges to formulate a unified optimization among Gaussian spheres within a neighborhood. Therefore, we predict smooth opacity values for these coupled Gaussian spheres, which serve as differentiable criteria for identifying and removing redundant elements. 
Rather than optimizing each Gaussian's opacity independently—which causes chaotic holes and floaters—we leverage tetrahedral edges for a spatially coherent opacity field, a differentiable criterion for pruning redundant elements that breaks the reliance on fixed initializations.
% Rather than optimizing the opacity of each Gaussian independently—which often leads to chaotic internal holes or floating artifacts—we leverage tetrahedral edges to formulate a spatially coherent opacity regularization. This continuous opacity field serves as a robust, differentiable criterion for identifying and dynamically pruning redundant elements, breaking the reliance on fixed initializations.
However, this pruning process creates a secondary challenge: when internal elements become surface elements after deletion, the bi-harmonic energy~\cite{botsch2007linear} optimization fails to smooth the surface effectively. To address this, we introduce an alternating geometric optimization strategy between tetrahedral elements and their coupled Gaussian spheres. Specifically, we implement a  Lagrangian smoothing approach that regularizes element positions while preserving global structure with a weighted bi-harmonic energy. To further enhance geometric quality and prevent surface shrinkage---a common artifact in mesh smoothing---we develop a novel two-stage smoothing strategy that significantly improves the performance of geometric optimization during reconstruction.

% Extensive experiments demonstrate that our method achieves superior reconstruction accuracy compared to state-of-the-art techniques across multiple benchmarks. Notably, our coupled representation design delivers substantial rendering quality improvements over alternative geometry reconstruction methods. Most importantly, by generating a single, watertight tetrahedral mesh representation through our holistic optimization framework, our reconstructions are inherently suitable for physical simulation without additional processing. We validate this capability through multiple simulation scenarios, achieving realistic physical behaviors that demonstrate our method's dual strengths in geometric precision and simulation readiness. Our key contributions are:

Extensive experiments demonstrate that our method achieves superior reconstruction accuracy compared to state-of-the-art techniques across multiple benchmarks. Notably, our coupled representation delivers substantial rendering quality improvements over alternative geometry reconstruction methods. Most importantly, by generating a unified and topologically coherent tetrahedral mesh through our holistic optimization framework, our reconstructions are suitable for physical simulation, avoiding the fragile conventional surface-extraction-and-tetrahedralization pipeline. We validate this capability through multiple simulation scenarios, achieving realistic physical behaviors that demonstrate our method's dual strengths in geometric precision and simulation compatibility. Our key contributions are:

\begin{itemize}
    \item We introduce a novel representation that tightly couples tetrahedral mesh elements with Gaussian spheres, enabling both high-quality rendering and topology-adaptive optimization within a holistic framework. 
    \item We develop a differentiable pruning mechanism based on a smooth opacity prediction, allowing the mesh to dynamically adapt its topology during optimization rather than relying on fixed initialization. 
    \item We propose an alternating optimization approach with a two-stage smoothing strategy that preserves surface integrity after topological changes, effectively addressing the geometric degradation that occurs when internal primitives become surface elements. 
\end{itemize}

\section{Related Works}
\label{sec:related}

\subsection{Eulerian vs. Lagrangian Representations}
Eulerian and Lagrangian formulations are two of the most commonly used 
geometric representations in computer graphics and physics~\cite{chung2002computational}. 
The Lagrangian perspective~\cite{yariv2024mosaic} follows individual geometric primitives as they move through space, while Eulerian representations~\cite{shen2021dmtet,yang2019tet} model geometry as fields defined over static spatial locations and recover shape properties by querying these fields. Neural implicit methods~\cite{park2019deepsdf,yariv2021volume,mildenhall2021nerf,wang2021neus,li2023neuralangelo} follow this paradigm by mapping 3D positions to signed distances or radiance values. These approaches offer continuous and potentially high-resolution geometry, but their dependence on neural network optimization often results in slow convergence and significant computational overhead. 
In contrast, Lagrangian approaches treat geometric elements---points, triangles, or volumetric cells---as entities that deform directly in space. Gaussian Splatting~\cite{kerbl20233d} is a prominent example, optimizing the positions and attributes of Gaussian primitives in 3D. Recent methods~\cite{son2024dmesh,shen2023flexible,gao2020learning,kulhanek2023tetra,gao2025curve,gao2025self}also adopt Lagrangian viewpoint by proposing an isosurface representation designed for optimizing unknown meshes through tetrahedral and hierarchically adaptive meshes. More recently, TetSphere Splatting~\cite{guo2024tetsphere} employs tetrahedral spheres as the fundamental volumetric primitive. The final reconstruction is represented as the union of multiple independent TetSpheres, where shape recovery is achieved by deforming each tetrahedral sphere individually. Despite achieving strong reconstruction results, this design relies on a collection of disjoint tetrahedral meshes, which limits its applicability to physical simulation and makes topology optimization difficult.

\subsection{Multi-view Reconstruction}  

Multi-view 3D reconstruction has long been a central topic in computer vision.
Classical pipelines rely on structure-from-motion (SfM)~\cite{schoenberger2016sfm} to estimate camera poses and multi-view stereo (MVS) to obtain dense geometric correspondences. Recent progress has shifted toward differentiable rendering–based approaches, which recover geometry by optimizing a 3D representation to match multi-view image observations.
NeRF~\cite{mildenhall2021nerf} and its variants~\cite{neus2,wang2021neus,10926495} demonstrate that continuous volumetric functions can be optimized through differentiable volumetric rendering.
More recently, point-based formulations such as 3D Gaussian Splatting~\cite{kerbl20233d} show that explicit primitives can achieve real-time rendering and strong view reconstruction, inspiring numerous extensions in surface extraction~\cite{guedon2024sugar, chen2023neusg}, dynamic modeling, and generalizable reconstruction.  Parallel to optimization-based methods, the field has also seen the rise of large feed-forward 3D reconstruction models.
Models such as the 3R family~\cite{yang2025fast3r,wang2024dust3r} and VGGT~\cite{wang2025vggt} learn to infer 3D geometry directly from images through large-scale pretraining and transformer architectures, significantly improving reconstruction quality and generalization.

% However, most existing methods~\cite{son2024dmesh, huang20242d, NeuS, son2024dmesh, chen2024meshanything} primarily focus on recovering surfaces and do not naturally yield volumetric structures suitable for downstream physical simulations. Our method reconstructs a holistic tetrahedral mesh that is both watertight and volumetric, providing a simulation-ready representation.

However, most existing methods~\cite{huang20242d, wang2021neus, son2024dmesh, chen2024meshanything} primarily focus on recovering surface geometry, inherently lacking the volumetric coherence required for downstream physical simulations. While recent advancements have integrated physical dynamics into neural representations~\cite{li2023pac, xie2024physgaussian,zhang2024physdreamer,huang2024dreamphysics}, these approaches predominantly rely on discrete particle-based proxies (e.g., Material Point Method), which struggle to preserve strict topological integrity for continuous solid mechanics. 
In contrast, our method directly reconstructs a holistic, unified and topologically coherent tetrahedral mesh, providing a robust geometric foundation highly amenable to physical simulators.

\section{Method}

% In this paper, we propose a topology-adaptive TetSphere reconstruction framework that achieves a holistic and watertight tetrahedral mesh, ready to use in various physical simulations. 
% We provide the problem statement in \cref{sec:problem}; then we introduce several essential preliminaries of TetSphere Splatting in \cref{sec:pre};
% \cref{sec:init} then describes the formulation of the first topology optimization component, where we get the topology-adaptive tetrahedral mesh;
% Subsequently, \cref{sec:opt} discusses the second optimization component, detailing our overall geometric refinement strategy. 

In this paper, we propose a topology-adaptive TetSphere reconstruction framework that achieves holistic, unified and topologically coherent tetrahedral meshes, providing a geometric foundation for physical simulations. 
After the problem statement (\cref{sec:problem}) and TetSphere preliminaries (\cref{sec:pre}), \cref{sec:init} details our topology optimization and \cref{sec:opt} the alternating geometric refinement that recovers surface detail while preserving topology.
\subsection{Problem Statement}\label{sec:problem}
TetSphere Splatting~\cite{guo2024tetsphere} optimizes the vertex coordinates $\{v\}$ across all tetrahedral spheres via a homeomorphism:
\vspace{-8pt}
\begin{equation}
f_\theta(v_i): \Omega \subset \mathbb{R}^3 \longrightarrow \mathbb{R}^3,
\quad
v_i' = f_\theta(v_i;\theta_v),
\label{eq:homeomorphism}
\vspace{-8pt}
\end{equation}

\noindent where $\theta_v$ denotes the learnable deformation parameters.  
However, the homeomorphism preserves the topology of the initial tetrahedral mesh, making it highly dependent on initialization and fundamentally limiting its ability to model shapes with complex topology. 

We aim to propose a topology-adaptive TetSphere reconstruction framework that achieves a holistic and watertight tetrahedral mesh through end-to-end topological and geometric optimization. The key challenge is to handle topology and geometry simultaneously.  

Thus, we formulate a joint optimization objective conceptually as:
\vspace{-5pt}
\begin{equation}
\min_{\Theta}
\; \mathcal{L}_{\text{conceptual}} =
\underbrace{
\mathcal{D}_{\text{topo}}(\hat{T}, T)
}_{\text{Topology}}
+
\underbrace{
\mathcal{D}_{\text{geom}}(\hat{G}, G)
}_{\text{Geometry}},
\label{eq:joint_loss}
\vspace{-5pt}
\end{equation}
where $\hat{T}$ and $T$ represent the predicted and target topologies, while $\hat{G}$ and $G$ denote the predicted and target geometries, respectively. $\mathcal{D}_{\text{topo}}$ and $\mathcal{D}_{\text{geom}}$ represent abstract discrepancy measures for topology and geometry, and $\Theta$ denotes all learnable parameters. 

It is worth noting that \cref{eq:joint_loss} serves as a high-level conceptual formulation. Since directly computing the gradient of discrete topological metrics is intractable, we introduce a differentiable continuous opacity field as a tractable surrogate in later sections to effectively approximate and solve this joint objective.

\subsection{Preliminaries} \label{sec:pre}

The key insight of  TetSphere Splatting~\cite{guo2024tetsphere}  is that conventional Lagrangian primitives (\eg, point clouds) are too fine-grained to ensure global mesh regularity. 
Instead, the TetSphere representation employs \emph{tetrahedral spheres} as volumetric primitives---each TetSphere is a volumetric unit composed of a set of vertices connected by tetrahedral elements. Specifically, the method first performs silhouette coverage~\cite{dou2022coverage} to select the initial centers of TetSpheres using feature points of the target, and initializes multiple tetrahedralized spheres $\{M\}$ represented as a set of tetrahedral elements $\{ m_i \}_{i=1}^{N}$, 
where each $m_i = (v_{i,1}, v_{i,2}, v_{i,3}, v_{i,4})$ denotes a tetrahedron defined by its four vertices.  

 Let $\mathbf{x}$ denotes the stacked vertex coordinates of all tetrahedral elements. The deformation of TetSpheres is obtained by minimizing the discrepancy between the rendered appearance and the input images.  
In addition to the rendering loss, the deformation field is regularized through a bi-harmonic~\cite{botsch2007linear} smoothness term defined on the deformation gradients $\mathbf{F}_{\mathbf{x}} = \mathbf{D}_s\mathbf{D}_m^{-1} $, controlled by a discrete Laplacian operator $\mathbf{L}$.   Let the undeformed tetrahedral vertices~\cite{schuller2013locally} be 
$\mathbf{X}=\{\mathbf{X}^{(i)}\}_{i=1}^{4}$ 
and the deformed vertices be 
$\mathbf{x}=\{\mathbf{x}^{(i)}\}_{i=1}^{4}$.  
% The matrices $\mathbf{D}_s$ and $\mathbf{D}_m$ are defined as:
% \begin{align}
% \mathbf{D}_s &:= 
% \left[\mathbf{x}^{(1)}-\mathbf{x}^{(4)},\;
%        \mathbf{x}^{(2)}-\mathbf{x}^{(4)},\;
%        \mathbf{x}^{(3)}-\mathbf{x}^{(4)}\right],\\
% \mathbf{D}_m &:=
% \left[\mathbf{X}^{(1)}-\mathbf{X}^{(4)},\;
%        \mathbf{X}^{(2)}-\mathbf{X}^{(4)},\;
%        \mathbf{X}^{(3)}-\mathbf{X}^{(4)}\right].
% \end{align}
The matrices $\mathbf{D}_s$ and $\mathbf{D}_m$ are defined exactly as in ~\cite{guo2024tetsphere}.

Moreover, each tetrahedron must preserve its orientation, requiring the determinant of its Jacobian $\mathrm{det}(\mathbf{F}^{(i,j)}_{\mathbf{x}})$~\cite{sifakis2012fem} to remain positive.  
Formally, the deformation is characterized by the constrained objective:
\vspace{-5pt}
% \begin{align}
% \min_{\mathbf{x}} \quad
% & \mathbf{\Phi}(R(\mathbf{x})) + \lVert \mathbf{L}\mathbf{F}_{\mathbf{x}} \rVert_2^2 \label{eq:tet_opt}, \\
% \text{s.t.} \quad
% & \mathrm{det}(\mathbf{F}^{(i,j)}_{\mathbf{x}}) > 0,\;
% \forall i \in \{1,\ldots,M\},\;
% j \in \{1,\ldots,T\}, \notag
% \vspace{-5pt}
% \end{align}

\vspace{-5pt}
\begin{equation}
\min_{\mathbf{x}} \; \mathbf{\Phi}(R(\mathbf{x})) + \lVert \mathbf{L}\mathbf{F}_{\mathbf{x}} \rVert_2^2, \quad \text{s.t.} \;\; \mathrm{det}(\mathbf{F}^{(i,j)}_{\mathbf{x}}) > 0, \;\; \forall i, j.
\label{eq:tet_opt}
\vspace{-5pt}
\end{equation}
where \noindent $R(\cdot)$ indicates a rendering operator and $\mathbf{\Phi}(\cdot)$ is a photometric objective.  However, such homeomorphic deformations inherently preserve topology and cannot efficiently accommodate topological variations, for example, the creation of hollow regions inside the object.  
In contrast, our method operates on a single hull-bounded tetrahedral volume with a shared vertex-opacity field and differentiable pruning, allowing the connectivity itself—not just vertex positions—to change during optimization.

\subsection{Tetrahedral Topology Optimization} \label{sec:init}

To address the limitation of homeomorphic deformation, we propose a
\emph{Topology-Adaptive Tetrahedral Representation} that enables
flexible topology modification by selectively pruning elements while
maintaining continuity.
We redefine the deformation domain to exclude pruned regions:
\vspace{-5pt}
\begin{equation}
\begin{aligned}
& \Omega_{\text{valid}} = \Omega \setminus D_\theta, \quad
  D_\theta = \{\, x \in \Omega \mid w_\theta(x) = 0 \,\}, \\%[3pt]
& f_\theta(v_i, \theta_v) : \Omega_{\text{valid}}
  \;\longrightarrow\; \mathbb{R}^3, \quad v_i' = f_\theta(v_i;\,\theta_v),
\end{aligned}
\label{eq:topo_varying_deform}
\vspace{-5pt}
\end{equation}
where $w_\theta(x)\!\in\![0,1]$ determines element existence,
$\Omega_{\text{valid}}$ is the pruned reference domain, and $\theta_v$
are deformation parameters, generalizing
Eq.~\eqref{eq:homeomorphism} to allow topology adaptation.

\vspace{2pt}\noindent\textbf{Robust Initialization.}
Rather than relying on sparse SfM features, we employ a two-stage
strategy:
(1) rapid 2DGS reconstruction~\cite{huang20242d} provides a coarse point cloud;
(2) convex-hull computation from Gaussian centers, followed by Delaunay
tetrahedralization, constructs a watertight initial mesh $M_0$.

\vspace{2pt}\noindent\textbf{Continuous Opacity Field.}
To identify redundant regions, we place a Gaussian in each
tetrahedron with opacity $\alpha_i$ indicating membership, inspired by continuous topology optimization~\cite{bendsoe2013topology,zehnder2021ntopo}.
Instead of independent per-element optimization (which causes
discontinuities), we assign each vertex a learnable scalar field value
$\phi$ (a \emph{vertex implicit field}) and derive element opacity via
barycentric mean followed by a sigmoid mapping:
$
\bar{\phi}_i = \frac{1}{4}\sum_{j=1}^{4}\phi_{ij}, \
\alpha_i = \sigma\!\left(-s \cdot \bar{\phi}_i \cdot \kappa\right),
\label{eq:opacity_derive}
$
where $\sigma$ is the sigmoid function, $s$ is a fixed scale factor,
and $\kappa$ is a learnable sharpness parameter that controls the transition steepness between inside and outside
regions. Note that $\phi$ itself is not directly the opacity but a
continuous implicit field from which opacity is derived.

\vspace{2pt}\noindent\textbf{Design Principle: Spatial Coherence through Vertex Sharing.}
Our vertex-shared parameterization is motivated by spatially coherent pruning: optimizing per-tetrahedron opacity independently empirically yields chaotic fragmentation, whereas barycentric averaging plus the smoothness regularizer $\mathcal{L}_{\text{smooth}}$ correlates adjacent elements' opacities, so pruning boundaries form smooth connected regions rather than scattered deletions.
% Our vertex-based opacity parameterization is motivated by the need for 
% spatially coherent pruning decisions. Rather than optimizing opacity 
% independently for each tetrahedron—which we empirically observe leads to 
% chaotic fragmentation—we design 
% a shared parameterization that encourages smooth opacity variation.

% \textbf{Spatial Coherence Properties.} This design has several desirable 
% properties that we verify empirically:

% \emph{(1) Bounded Opacity Variation:} For adjacent tetrahedra $t_i, t_j$ 
% sharing vertices, their opacities $\alpha_i, \alpha_j$ tend to be similar. 
% Intuitively, the barycentric averaging creates spatial correlation, and 
% the smoothness regularization $\mathcal{L}_{\text{smooth}}$
% further limits abrupt changes between neighboring elements.

% \emph{(2) Coherent Pruning Boundaries:} Empirically, we observe that 
% pruning boundaries form smooth, connected regions rather than scattered 
% deletions. This occurs because: shared vertex weights create local correlation, and our regularization discourages high-frequency weight variations.

We validate these properties through ablation (Tab. \ref{table:ablation}): removing the 
continuous opacity field (\textit{w/o Con-opacity}) fragments the mesh 
into 133 components, while our full method maintains connectivity 
(96.7\% single-component rate).

% \vspace{2pt}
% The final topology optimization via:
% \vspace{-5pt}
% \begin{equation}
% \min_{\{\phi\}} \mathcal{L}_{\text{topo}} =
% \mathcal{L}_{\text{render}}
% + \lambda_{\text{eik}}  \mathcal{L}_{\text{eik}}(\nabla\Psi)
% + \lambda_{\text{smooth}} \mathcal{L}_{\text{smooth}}(\phi),
% \label{eq:topo_loss}
% \vspace{-5pt}
% \end{equation}
% where $\mathcal{L}_{\text{render}}$ aligns the reconstructed shape with
% multi-view observations,
% $\mathcal{L}_{\text{eik}}$ penalises excessively steep gradients of
% $\Psi$ to avoid abrupt opacity transitions and
% $\mathcal{L}_{\text{smooth}}$ regularizes $\phi$ to prevent isolated
% low-opacity tetrahedra.

The final topology optimization objective is:
\vspace{-6pt}
\begin{equation}
\min_{\{\phi\}} \mathcal{L}_{\mathrm{topo}} = \mathcal{L}_{\mathrm{render}} + 
\lambda_{\mathrm{eik}} \mathcal{L}_{\mathrm{eik}} + \lambda_{\mathrm{smooth}} 
\mathcal{L}_{\mathrm{smooth}},
\vspace{-6pt}
\end{equation}
where $\mathcal{L}_{\mathrm{render}}$ aligns the reconstructed shape with multi-view 
observations. The regularization terms are defined over the mesh edge set $\mathcal{E}$:
\vspace{-6pt}
\begin{equation}
\mathcal{L}_{\mathrm{eik}} = \frac{1}{|\mathcal{E}|} \sum_{(u,v) \in \mathcal{E}} 
\left( \frac{|\phi_u - \phi_v|}{\ell_{uv}} - 1 \right)^2, \quad
\mathcal{L}_{\mathrm{smooth}} = \frac{1}{|\mathcal{E}|} \sum_{(u,v) \in \mathcal{E}} 
(\phi_u - \phi_v)^2,
\vspace{-6pt}
\end{equation}
where $\ell_{uv} = \|x_u - x_v\|_2$ is the edge length. 
$\mathcal{L}_{\mathrm{eik}}$ constrains the
\emph{rate of change} of $\phi$, preventing the field from
collapsing or exploding spatially.
$\mathcal{L}_{\mathrm{smooth}}$ penalizes absolute
adjacent differences to suppress local oscillations.
The two are complementary: the former governs global variation rate,
the latter controls local high-frequency noise.

\subsection{Tetrahedral Geometry Reconstruction} \label{sec:opt}
Here, we describe the alternating optimization framework together with a two-stage smoothing strategy for stabilizing tetrahedral geometry optimization after topological updates.

\vspace{2pt}\noindent\textbf{Alternating optimization}
We alternate between optimizing the tetrahedra and the Gaussians.  
Within our hybrid representation, updating the tetrahedral elements naturally induces the motion of the Gaussians attached to them.  
Formally, the center of each Gaussian $\mathbf{g}_k$ is updated according to the barycentric interpolation of the tetrahedral vertices as $\mathbf{g}_k = \sum_{i=1}^{4} \beta_{ik}\,\mathbf{v}_i,$
where $\mathbf{v}_i$ denote the tetrahedral vertex positions. Conversely, Gaussians are rapidly optimized through image-based rendering losses, and their resulting surface normals $\mathbf{n}^{\text{GS}}$ supervise the rasterized normals of the tetrahedral mesh $\mathbf{n}^{\text{Tet}}$, guiding the refinement of tetrahedral geometry,
\(
\mathbf{n}^{\text{GS}}_k \;\longrightarrow\; \mathbf{n}^{\text{Tet}}_i.
\)
This normal-based supervision flow from Gaussians to tetrahedral elements is practically enforced via a masked cosine distance loss: $\mathcal{L}_{\text{norm}} = \sum_{p} M_{\text{mask}}(p) \big(1 - \langle \mathbf{n}^{\text{GS}}_p, \mathbf{n}^{\text{Tet}}_p \rangle\big)$, where $M_{\text{mask}}(p)$ is a binary mask
that selects pixels with non-zero rendered alpha, restricting the
normal supervision to visible surface regions.

\vspace{2pt}\noindent\textbf{Two-stage Laplacian Smoothing.}
The tetrahedral mesh is regularized through a two-stage smoothing strategy that combines a weighted bi-harmonic energy with a Laplacian-based term, ensuring stable and high-quality geometric updates throughout optimization. Based on the observation that the vanilla Laplacian energy often causes global shrinkage of the tetrahedral mesh~\cite{desbrun1999implicit,taubin1995signal}, we utilize a normal-decomposed \emph{HC-Laplacian} energy~\cite{Vollmer1999ImprovedLS} that separates vertex displacements into tangential and normal components. 
This energy is optimized in two stages: the first stage focuses on pushing out concave regions, and the second emphasizes overall surface smoothing, leading to a refined and continuous final geometry. The formulation of the filtered HC-Laplacian energy over surface vertices $\mathcal{S}$ is defined as:
\vspace{-5pt}
\begin{equation}
\mathcal{L}_{\text{HC}} 
= \sum_{i \in \mathcal{S}} 
\left[
\|d_i^{t}\|^{2} 
+ \lambda_{\text{cond}}(t)\,(d_i^{n})_{-}^{2} 
+ (d_i^{n})_{+}^{2}
\right],
\vspace{-5pt}
\end{equation}
where the surface vertex set $\mathcal{S}$ is strictly defined by vertices belonging to the boundary faces $\mathcal{F}_{\partial} = \{ f \mid \text{count}(f) = 1 \}$; $d_i^{t}$ and $d_i^{n}$ denote the tangential and normal components of the vertex displacement; $(d_i^{n})_{+}$ and  $(d_i^{n})_{-}$ denote the outward and inward normal displacements of the tetrahedra, respectively. The stage-dependent weight $\lambda_{\text{cond}}(t)$ is set to $0$ for Stage 1: concavity pushing and $1$ for Stage 2: global smoothing.

% A standard bi-harmonic energy is adopted by TetSphere Splatting~\cite{guo2024tetsphere} to penalize higher-order variations of vertex motion and thus preserves the smooth characteristics of the mesh during deformation, as formulated in \cref{eq:tet_opt}. TetSphere Splatting benefits from this regularization because it initializes from smooth TetSpheres. In contrast, our framework begins with a structure that may include uneven surfaces or high-frequency geometric variations. In such cases, applying the standard bi-harmonic energy uniformly can excessively damp local deformations, preventing accurate reconstruction of fine geometric details.

TetSphere ~\cite{guo2024tetsphere} initializes from smooth TetSpheres and thus benefits from a standard bi-harmonic energy \cref{eq:tet_opt}. Our framework instead starts from uneven, high-frequency geometry, where applying it uniformly overly damps local deformations and loses fine detail.

To balance global smoothness with local geometric fidelity, we introduce a \emph{weighted bi-harmonic energy}, which reduces the smoothness constraint specifically for surface or high-detailed regions:
% \begin{equation}
% \mathcal{L}_{\text{weighted}} = w(\mathbf{x}) \, \lVert \mathbf{L}\mathbf{F}_{\mathbf{x}} \rVert_2^2, \quad \text{with } \;
% w(\mathbf{x}_i) =
% \begin{cases}
% \gamma < 1, & \text{if } \mathbf{x}_i \in \mathcal{S}, \\
% 1, & \text{otherwise}.
% \end{cases}
% \label{eq:weighted_energy}
% \vspace{-10pt}
% \end{equation}
\vspace{-5pt}
\begin{equation}
\mathcal{L}_{\text{w}} = w(\mathbf{x}) \, \lVert \mathbf{L}\mathbf{F}_{\mathbf{x}} \rVert_2^2, \quad \text{with } \; w(\mathbf{x}_i) = \gamma \; (\gamma < 1) \text{ if } \mathbf{x}_i \in \mathcal{S}, \text{ and } 1 \text{ otherwise}.
\label{eq:weighted_energy}
\vspace{-5pt}
\end{equation}
Here $\mathcal{S}$ is identical to the surface vertex set in
$\mathcal{L}_{\mathrm{HC}}$ and is recomputed after each topological update.
This design gives the surface vertices ($\mathcal{S}$) greater freedom of deformation to fit fine geometric details, while the interior vertices remain strongly regularized to ensure structural stability. 

Ultimately, the joint geometry optimization is guided by a combination of image-based silhouette supervision and the structurally preserving regularization:
\vspace{-5pt}
\begin{equation}
  \mathcal{L}_{\text{geo}}
  = \lambda_{\text{m}}  \,\mathcal{L}_{\text{mask}}
  + \lambda_{\text{n}}  \,\mathcal{L}_{\text{norm}}
  + \lambda_{\text{HC}}  \,\mathcal{L}_{\text{HC}}
  + \lambda_{\text{w}}  \,\mathcal{L}_{\text{w}}.
  \label{eq:geo_loss}
  \vspace{-5pt}
\end{equation}

Where the $\mathcal{L}_{\text{mask}}$ penalizes discrepancies between the rendered
opacity map $\hat{O}$ and the foreground silhouette mask $M_{\text{gt}}$ as $
  \mathcal{L}_{\text{mask}}
  = \bigl\|\hat{O} - M_{\text{gt}}\bigr\|_1
  \label{eq:lmask}
$.

With the above regularization loss and rendering loss, our coupled geometry optimization effectively produces a unified, topologically coherent tetrahedral mesh with high geometric accuracy.

\section{Experiments} \label{sec:experiments}

\subsection{Experiment Settings}

\textbf{Datasets.} 
%Following the experimental protocol of TetSphere Splatting~\cite{guo2024tetsphere}, w
We evaluate on a combined 3D dataset containing objects that span both closed- and open-surface geometries. Specifically, we use eight watertight models from the Thingi10k collection~\cite{zhou2016thingi10k}, four open-surface objects from the DeepFashion3D dataset~\cite{zhu2020deep}, eight watertight models from Objaverse~\cite{deitke2023objaverse}, and ten objects from the Google Scanned Objects dataset~\cite{GSO}. Each scene provides 120 input images, and we select every 10th frame as test viewpoints with the remaining for training.

\noindent \textbf{Metrics.} We adopt standard surface metrics including \emph{Chamfer $L_1/L_2$, Recall, Volume IoU}, and \emph{Hausdorff distance}. For ablation studies, we further report \emph{F-Score, Normal Consistency}, \emph{Edge Precision/F-Score}, and \emph{Number of tetrahedral components (Num. components)}. Rendering quality is measured via \emph{PSNR, SSIM}, and \emph{LPIPS}. To assess tetrahedral mesh quality and FEM stability, we evaluate \emph{Minimum Dihedral Angle, Aspect Ratio, Inverted Ratio}, and \emph{Manifoldness Rate}. Additionally, we report \emph{Single/Multiple Component ratios} and \emph{Average Component number} to quantify topological connectivity against TetGen~\cite{hang2015tetgen}-based baselines (e.g., NeuS2, 2DGS, DMesh).

\noindent \textbf{Baselines.} 
For geometric comparison, we quantitatively evaluate our method against a broad set of state-of-the-art approaches spanning both Eulerian and Lagrangian representations.  
The Eulerian category includes NeuS2~\cite{neus2}.  
The Lagrangian category includes three representative explicit or hybrid approaches: 2DGS~\cite{huang20242d}, DMesh~\cite{son2024dmesh}, and TetSphere Splatting~\cite{guo2024tetsphere}.

\noindent\textbf{Implementation Details.} 
We implement HoloTetSphere on a single NVIDIA RTX 4090 (24 GB), optimizing for 30k iterations. Starting at 15k iterations, we adopt the topology optimization and an alternating schedule, switching between Gaussian and tetrahedral mesh optimization in 5k-iteration blocks. Please refer to the supplementary material for comprehensive hyperparameters and schedules.

% \begin{figure}[t]
%   \centering
%   % 左侧：表格垂直居中
%   \begin{minipage}[c]{0.48\linewidth}
%     \centering
%     \includegraphics[width=\linewidth]{figs/midcut.pdf}
%     \captionof{figure}{
%   Tetrahedra visualization. Unlike TetSphere's overlapping and disjoint primitives, our method produces  a single, integrated, and topologically coherent tetrahedral mesh.
%  }
%     \label{fig:midcut}
%   \end{minipage}
%   \hfill
%   % 右侧：图片垂直居中
%     \begin{minipage}[c]{0.48\linewidth}
%     \centering
%     \includegraphics[width=\linewidth]{figs/trainingprocess.pdf}
%     \captionof{figure}{
% PSNR comparison with 2DGS. Better geometry recovery helps our method surpass 2DGS in rendering.
%  }
%     \label{fig:trainingprocess}
%   \end{minipage}
%   \vspace{-20pt}
% \end{figure}
%We implement our HoloTetSphere framework using PyTorch. The optimization spans a total of 30,000 iterations, with alternating geometric and topological updates dynamically scheduled to ensure stable tetrahedral refinement as shown in \cref{fig:trainingprocess}. The continuous scalar field $\phi$ is implicitly modeled as a heuristic Signed Distance Function (SDF) to guide the differentiable topology pruning robustly. For comprehensive details regarding hyperparameter settings, initialization configurations, exact loss weights, and optimization schedules, please refer to Section 2 of our Supplementary Material.
\begin{figure}[t]
  \centering
  \includegraphics[width=\linewidth]{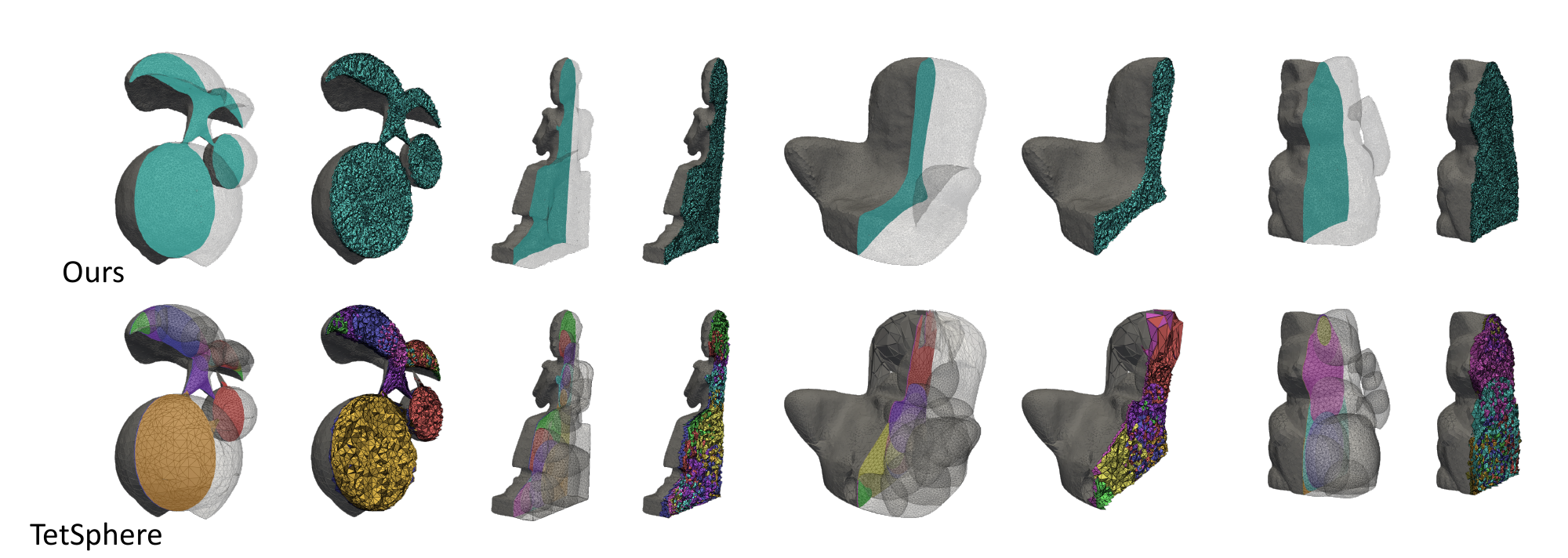}
  \vspace{-15pt}
 \caption{Tetrahedra visualization. For each object, the first column displays the whole structure, while the second column visualizes the interior tetrahedral wireframe. Different colors indicate distinct tetrahedral mesh components. Unlike TetSphere's overlapping and disjoint primitives, our method produces a single, integrated, and topologically coherent tetrahedral mesh.
 }
  \label{fig:midcut}
  \vspace{-20pt}
\end{figure}

\subsection{Geometry Comparison}

% \cref{table:comparison_extended} presents the quantitative comparison results.  
% Compared with all baselines, our method achieves superior reconstruction accuracy in terms of Chamfer Distance and Volume IoU.   Moreover, our approach produces uniform meshes comparable to TetSphere Splatting, as reflected by the low ALR, while effectively avoiding floating artifacts, as indicated by the reduced CC Diff. Importantly, HoloTetSphere consistently reconstructs tetrahedra forming a single coherent volumetric component, as captured by the CT metric, \ie, its mesh can support physical simulations directly without additional tetrahedralization. 

\cref{table:comparison_extended} presents the quantitative comparison results.  
Compared with all baselines, our method achieves superior overall reconstruction accuracy, yielding the best performance across all distance-based metrics, including Chamfer Distance, Chamfer L2, and Hausdorff distance. 
The markedly lower Hausdorff and Chamfer L2 indicate strong suppression of the structural outliers and floating artifacts common to other Lagrangian methods. Our method also stays competitive on completeness and volume, taking second-best Recall and Volume IoU—just behind NeuS2 and ahead of all other Lagrangian methods.
% Specifically, the significantly lower Hausdorff distance and Chamfer L2 demonstrate that our approach effectively suppresses large structural outliers and floating artifacts that often plague other Lagrangian methods. Furthermore, our approach maintains highly competitive completeness and volumetric accuracy, securing the second-best Recall and Volume IoU, closely trailing the Eulerian-based NeuS2 while significantly outperforming all other Lagrangian representations. 
Importantly, achieving such high geometric fidelity does not compromise our volumetric representation; HoloTetSphere directly reconstructs coherent tetrahedral meshes that serve as a volumetric basis for physical simulation (Sec.~\ref{sec:physim}), avoiding the fragile surface-extraction-and-tetrahedralization pipeline.

\definecolor{myyellow}{rgb}{1,1, 0.6}
\definecolor{myred}{rgb}{1, 0.6, 0.6}
\begin{table}[h!]
\setlength{\tabcolsep}{4.pt} % 进一步缩小列间距以容纳更多列
% \vspace{-10px}
\begin{center}
\vspace{-10pt}
\caption{Comparison of surface mesh reconstruction accuracy. Our method achieves the best distance-based metrics—highlighting its robustness against outliers and floating artifacts—while maintaining highly competitive volumetric completeness. We color code the \colorbox{myred}{\textbf{best}} and \colorbox{myyellow}{\textbf{second best}} results. 
}
\vspace{-10pt}
\label{table:comparison_extended}
\small
\resizebox{\linewidth}{!}{
\begin{tabular}{llcccccc}
\toprule
Method & Geo. Rep. & Rendering & Cham. ↓ & Cham. L2 ↓ & Rec. ↑ & Haus. ↓ & Vol. IoU ↑ \\
\midrule

NeuS2 & Eulerian & {\color{green}{\ding{51}}} & \colorbox{myyellow}{\textbf{0.0118}} & 0.0209 & \colorbox{myred}{\textbf{0.7754}} & 0.2419 & \colorbox{myred}{\textbf{0.7275}} \\

2DGS & Lagrangian & {\color{green}{\ding{51}}} & 0.0130 & \colorbox{myyellow}{\textbf{0.0190}} & 0.6678 & 0.1337 & 0.6557 \\
DMesh & Lagrangian & {\color{red}{\ding{55}}} & 0.0152 & 0.0213 & 0.6720 & \colorbox{myyellow}{\textbf{0.1173}} & 0.6572 \\
TetSphere & Lagrangian & {\color{green}{\ding{51}}} & 0.0173 & 0.0291 & 0.6644 & 0.1784 & 0.5712 \\
\midrule
\textbf{Ours} & Lagrangian & {\color{green}{\ding{51}}} & \colorbox{myred}{\textbf{0.0102}} & \colorbox{myred}{\textbf{0.0163}} & \colorbox{myyellow}{\textbf{0.7442}} & \colorbox{myred}{\textbf{0.1118}} & \colorbox{myyellow}{\textbf{0.7268}} \\
\bottomrule
\end{tabular}
}
\end{center}
\vspace{-10pt}
\end{table}

\begin{table}[t]
\centering
\caption{Comparison of tetrahedral meshes. (Top) Tetrahedra reconstruction quality. (Bottom) Suitability for single-object physical simulation. Vol. denotes volumetric reconstruction; "-" indicates no additional tetrahedralization is required. A valid object should form a single connected mesh; thus, a higher Single-Comp. rate and Avg. Comp. approaching 1 indicate better suitability for physical simulation.}
\label{table:two_tab}
\vspace{-2mm} % 压缩表头与表格的间距
\footnotesize % 缩小一号字体，适配双表合并后的版面

% 第一部分：原来的 Table 2
\begin{tabular}{lcccc}
\toprule
Method & Min Dihedral $\uparrow$ & Aspect Ratio $\downarrow$ & Inverted Ratio $\downarrow$ & MR $\uparrow$ \\
\midrule
TetSphere & 32.000 & 2.587 & 0.057 & \textbf{100.0\%} \\
\textbf{Ours} & \textbf{39.503} & \textbf{2.554} & \textbf{0.017} & {96.7\%} \\
\bottomrule
\end{tabular}

\vspace{3mm} % 两个子表中间留出清晰的分界缝隙

% 第二部分：原来的 Table 3
\begin{tabular}{lccccc}
\toprule
Methods & Vol. & Single-Comp. $\uparrow$ & Multi-Comp. $\downarrow$ & Avg. Comp. $\downarrow$ & Failed Ratio $\downarrow$ \\
\midrule
NeuS2     & {\color{red}{\ding{55}}} & 60.0\% & 30.0\% & 1.48  & 10.0\% \\
2DGS      & {\color{red}{\ding{55}}} & 40.0\% & 26.7\% & 2.10  & 33.3\% \\
DMesh     & {\color{red}{\ding{55}}} & 53.3\% & 46.7\% & 7.60  & 0.0\%  \\
TetSphere & {\color{green}{\ding{51}}} & 0.0\%  & 100\%  & 61.21 & -  \\
\midrule
\textbf{Ours} & {\color{green}{\ding{51}}} & \textbf{96.7\%} & \textbf{3.3\%} & \textbf{1.03}  & \textbf{-}  \\
\bottomrule
\end{tabular}
\vspace{-4mm} % 减少表格底部的多余空白，为正文抢空间
\end{table}

\begin{figure}[t]
  \centering
  \includegraphics[width=\linewidth]{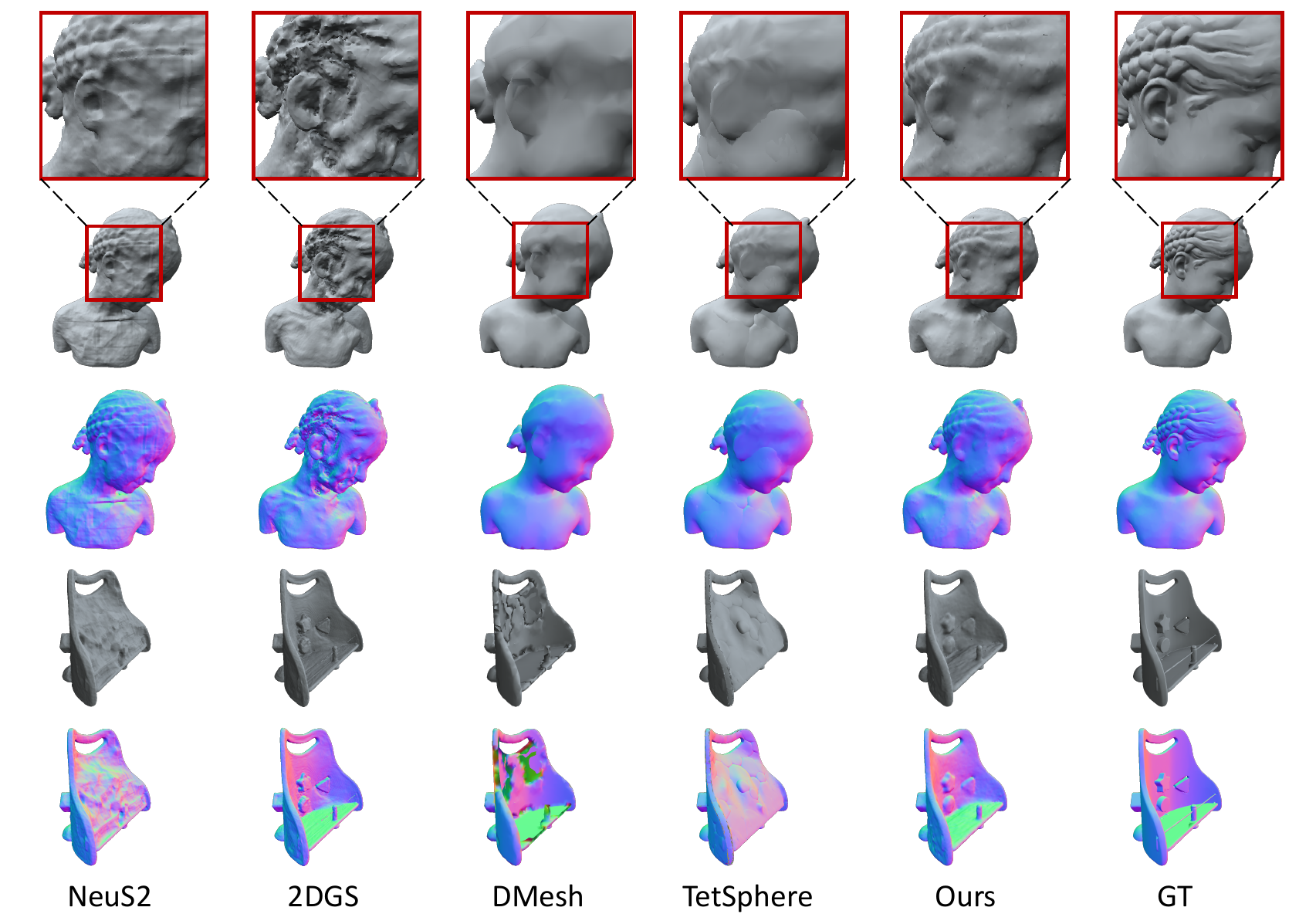}
  \vspace{-15pt}
 \caption{Qualitative comparisons of surface meshes. For each object, the top row shows the reconstructed mesh (with close-up views), and the bottom row visualizes the corresponding normal map. Our method achieves better connectivity, smoother geometry, and more accurate surface orientations.
 }
  \label{fig:geo_com}
  \vspace{-20pt}
\end{figure}

\cref{fig:geo_com} shows qualitative results across four representative shapes.  
Our method recovers richer geometric details.  
This improvement is largely attributed to the proposed adaptive topology optimization. As shown in the first object, TetSphere struggles to capture complex fine structures, 2DGS fails on textureless regions, and DMesh produces over-smoothed surfaces. In contrast, our method preserves clean and detailed geometry. More results are provided in the supplementary material.

\subsection{Physical Simulations}\label{sec:physim}

To validate our framework, we compare against surface-based methods (NeuS2, 2DGS, DMesh) combined with TetGen, and the direct volumetric method TetSphere. As shown in Table \ref{table:two_tab} (Bottom), traditional pipelines frequently fail during tetrahedralization or severely fragment due to non-watertight intermediate surfaces. While TetSphere bypasses this, it typically yields disjoint clusters. Conversely, our method achieves the highest rate in generating single, connected tetrahedral meshes, demonstrating superior topological integrity.

\begin{figure}[t]
  \centering
  % 左侧：表格垂直居中
  \begin{minipage}[c]{0.48\linewidth}
    \centering
    
    \small
    \captionof{table}{Deformation error comparison of physical simulation shown in \cref{fig:vis}. Sim. CD indicates the Chamfer distance between our simulated mesh and GT.}
    \label{table:isaac}
    \resizebox{\linewidth}{!}{
      \begin{tabular}{lccccc}
      \toprule
      \textbf{Time} & \textbf{$t{=}0.0$} & \textbf{$t{=}0.25$} & \textbf{$t{=}0.5$} & \textbf{$t{=}0.75$} & \textbf{$t{=}1.0$} \\
\midrule
\textbf{Sim. CD} & 0.0096 & 0.0119 & 0.0136 & 0.0171 & 0.0184 \\
\bottomrule
      \end{tabular}
    }
    
  \end{minipage}
  \hfill
  % 右侧：图片垂直居中
  \begin{minipage}[c]{0.48\linewidth}
    \centering
    \includegraphics[width=\linewidth]{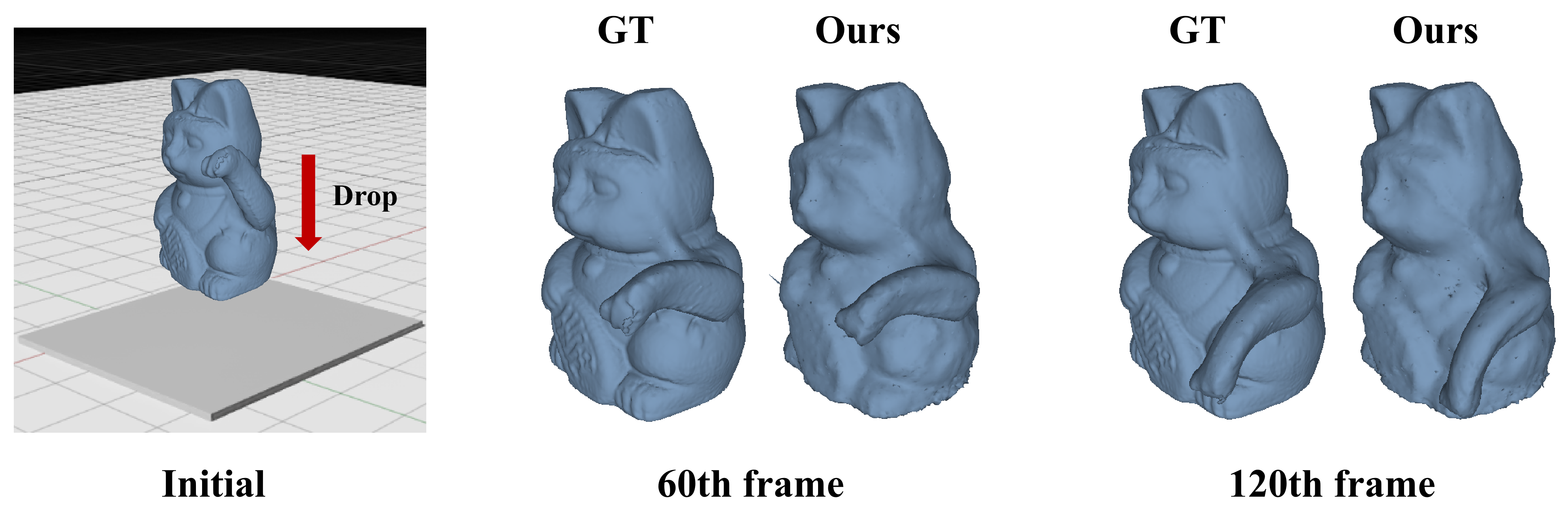}
    \caption{Visualization comparison of drop with deformation.} 
    \label{fig:vis}
  \end{minipage}
  \vspace{-10pt}
\end{figure}

\begin{figure}[t]
  \centering
  \includegraphics[width=\linewidth]{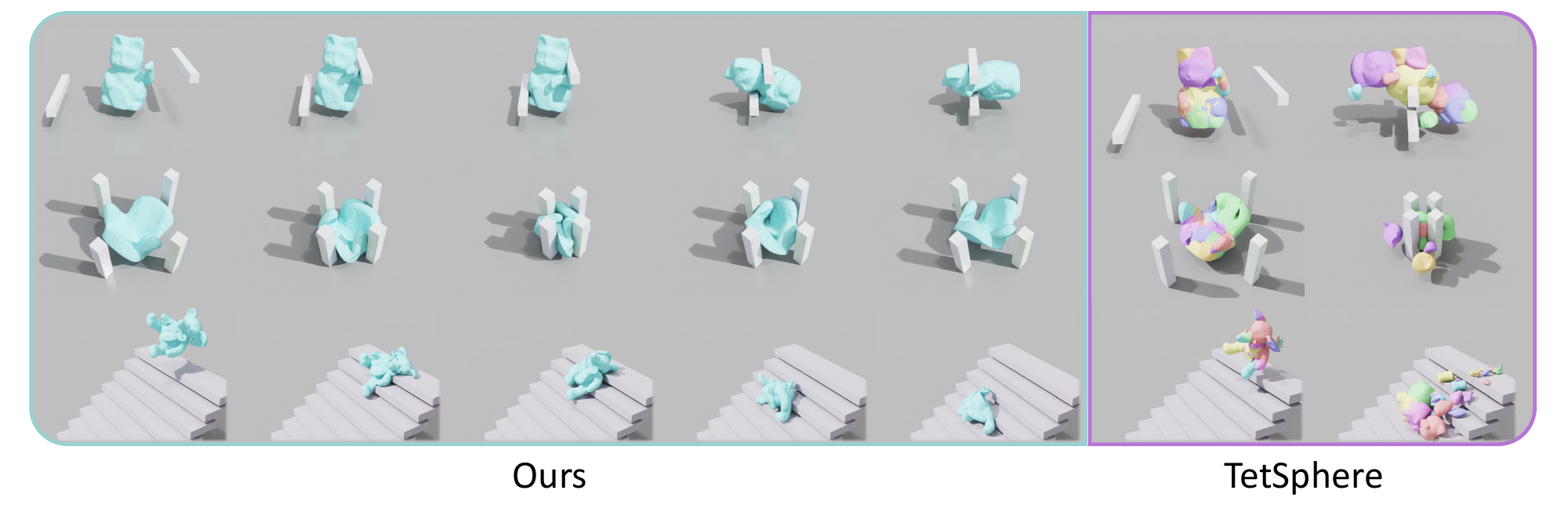}
  \vspace{-15pt}
 \caption{Physical simulation results. Our method achieves stable and realistic deformations, maintaining structural integrity under stress. In contrast, TetSphere fractures into multiple pieces, where different colors represent distinct, disconnected tetrahedral components.
 }
  \label{fig:physim}
  \vspace{-15pt}
\end{figure}

\cref{table:two_tab} (Top) assesses physical simulation suitability. Our method produces more regular elements with a higher Min Dihedral angle and lower Aspect Ratio, while significantly reducing the Inverted Ratio from 0.057 to 0.017. We stress that this 1.7\% is an upper bound on problematic elements: most inverted tetrahedra have neither face penetration nor degenerate volume and are restored simply by reordering their vertex indices, leaving only 0.14\% genuinely degenerate—and mostly surface-adjacent—elements, which a lightweight volume-threshold filter removes with negligible geometric error. After this automatic cleanup the meshes are inversion-free and numerically stable under FEM. Although TetSphere inherently maintains a 1.000 Manifoldness Rate via strict homeomorphic constraints, our approach achieves a robust 0.967 despite dynamic element pruning—which, after the cleanup above, suffices for stable FEM computations, effectively preserving topological integrity while enhancing overall geometric quality. Consequently, unlike TetSphere's disconnected components which trigger simulation instabilities (\cref{fig:midcut,fig:physim}), our unified, topologically coherent meshes support stable and realistic deformations. Beyond mesh-quality statistics, gravity-driven drop simulations against Isaac Sim FEM references show that the surface Chamfer distance stays low (below 0.02) and grows only gradually over the trajectory~(\cref{table:isaac,fig:vis}), confirming faithful surface-level deformation for uniform solid media. Simulation videos will be provided in supplementary.

% Finally, we evaluate scene-level multi-object reconstruction. By initializing scenes with a unified convex hull enclosing all objects, we reconstruct holistic meshes and segment individual instances via clustering for downstream simulations. As illustrated in \cref{fig:physical_sim}, this pipeline enables both stable physical simulations and subsequent physically-based rendering (we also provide supplementary video).

% \begin{figure}[t]
%     \centering
%     % 左侧子图：分配 0.58 的宽度（占多一点）
%     \begin{subfigure}{0.55\linewidth}
%         \centering
%         \includegraphics[width=\linewidth]{figs/tet_simulation.pdf}
%         \caption{Scene-level physical simulation.}
%         \label{fig:physical_sim}
%     \end{subfigure}
%     \hfill % 自动填充中间压力，将两图推向两端
%     % 右侧子图：分配 0.38 的宽度（占少一点）
%     \begin{subfigure}{0.42\linewidth}
%         \centering
%         \includegraphics[width=\linewidth]{figs/physim.pdf}
%         \caption{Compression Simulation.}
%         \label{fig:physim}
%     \end{subfigure}
%     \vspace{-5pt}
%     \caption{Physical simulation results. Left: Scene initialization encloses multiple objects within a unified convex hull, followed by clustering-based segmentation. Right: Our holistic mesh maintains stability under stress, whereas TetSphere fractures due to disjoint primitives.}
%     \label{fig:combined_results}
%     \vspace{-20pt}
% \end{figure}

\subsection{View-synthesis Comparison}
\begin{figure}[t]
  \centering
  % 左侧：表格垂直居中
  \begin{minipage}[c]{0.48\linewidth}
    \centering
    
    \small
    \captionof{table}{Quantitative comparison on the combined 3D dataset. Our topological pruning and optimization refine the geometry to surpass 2DGS, which struggles with textureless regions as shown in \cref{fig:geo_com}}.
    \label{table:tet_render}
    \resizebox{\linewidth}{!}{
      \begin{tabular}{lcccc}
      \toprule
           & NeuS2 & 2DGS & TetSphere & \textbf{Ours} \\
      \midrule
      PSNR $\uparrow$  & 37.73 & 40.25 & 34.75 & \textbf{40.68} \\
      SSIM $\uparrow$  & 0.986 & 0.985 & 0.977 & \textbf{0.987} \\
      LPIPS $\downarrow$ & 0.033 & 0.035 & 0.045 & \textbf{0.024} \\
      \bottomrule
      \end{tabular}
    }
    
  \end{minipage}
  \hfill
  % 右侧：图片垂直居中
  \begin{minipage}[c]{0.48\linewidth}
    \centering
    \includegraphics[width=\linewidth]{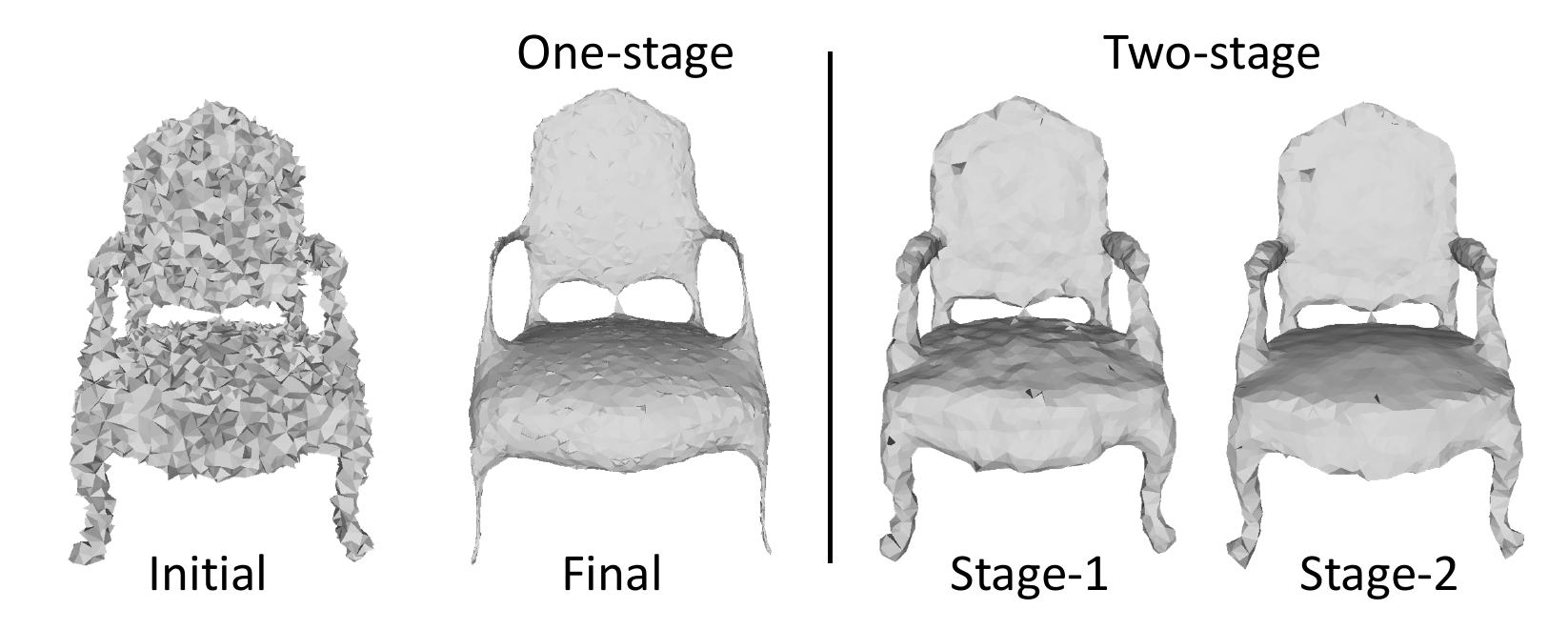}
    \captionof{figure}{Two-stage HC-Laplacian smoothing. Compared to the severe shrinkage in one-stage smoothing, our two-stage approach effectively preserves the original volume.}
    \label{fig:ablation_stage}
  \end{minipage}
  \vspace{-10pt}
\end{figure}

To evaluate the view-synthesis capability of our method, we conduct experiments on the combined 3D object dataset, as illustrated in \cref{table:tet_render}.
Benefiting from the joint optimization of tetrahedra and Gaussians, our method inherits the strong rendering capability of Gaussian splatting, leading to higher view-synthesis quality compared with other approaches.  
% To evaluate the efficiency performance, we conducted a running time comparison with TetSphere under optimal rendering settings. Our method achieves 50 FPS with a GPU memory consumption of 7.56 GB, whereas TetSphere achieves 30.76 FPS with 3.74 GB.
Under optimal rendering settings, our method runs at 50 FPS using 7.56 GB, versus TetSphere's 30.76 FPS at 3.74 GB.

\begin{wrapfigure}{r}{0.5\textwidth}
  \vspace{-15pt} % 可选：微调图片上方的空白距离
  \centering
  \includegraphics[width=\linewidth]{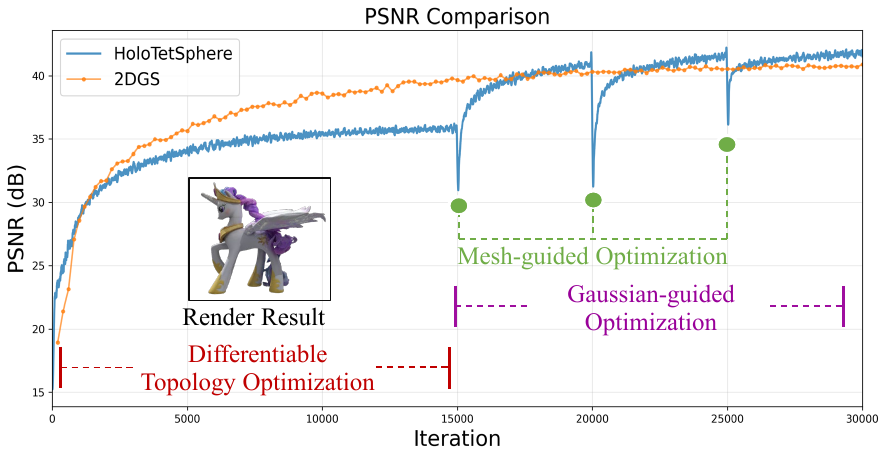}
  \caption{PSNR comparison with 2DGS. Better geometry recovery helps our method surpass 2DGS in rendering.}
  \label{fig:trainingprocess}
  \vspace{-10pt} % 可选：微调图片下方的空白距离
\end{wrapfigure}
In addition, the geometric consistency enforced by our tetrahedral mesh provides more stable and accurate Gaussian placements.
As shown in \cref{fig:trainingprocess}, we observe that our convex hull initialization provides a better starting point for rendering than 2DGS. While 2DGS gradually overtakes our method during the training process, the subsequent topological pruning and optimizing step eliminates redundant tetrahedra, refines our geometry and rendering quality, enabling our final results to surpass 2DGS. More results are in the supplementary.
%Our method outperforms 2DGS~\cite{huang20242d}, whose performance degrades on textureless regions and results in noticeably lower rendering quality under novel viewpoints. More results are in the supplementary.

\definecolor{myyellow}{rgb}{1,1, 0.6}
\definecolor{myorange}{rgb}{1, 0.8, 0.6}
\definecolor{myred}{rgb}{1, 0.6, 0.6}

% \begin{table}[t]

% \label{table:tet_render}
% \setlength{\tabcolsep}{3pt}
% \begin{center}
% \resizebox{0.85\linewidth}{!}{
% \begin{tabular}{lcccc}
% \toprule
%      &
%  \multicolumn{1}{c}{~~~NeuS2~ ~ ~ }&
%  \multicolumn{1}{c}{~~~ 2DGS~~~}&
%  \multicolumn{1}{c}{~~~Tetsphere~~~}&
%  \multicolumn{1}{c}{{\textbf{~~~Ours~~~}}}\\
% \midrule
% PSNR  $ \uparrow$ &37.73 & 40.25 & 34.75 & \textbf{40.68} \\
% SSIM  $ \uparrow$  & 0.986 & 0.985 & 0.977 & \textbf{0.987} \\
% LPIPS  $\downarrow$  & 0.033 & 0.035 & 0.045& \textbf{0.024}  \\ 
% \bottomrule
% \end{tabular}
% }

% \end{center}
% \caption{The quantitative comparison on the combined 3D dataset. We  highlight the best results. 
% }
% \end{table}

\subsection{Ablations}
\begin{figure}[t]
  \centering
  \includegraphics[width=\linewidth]{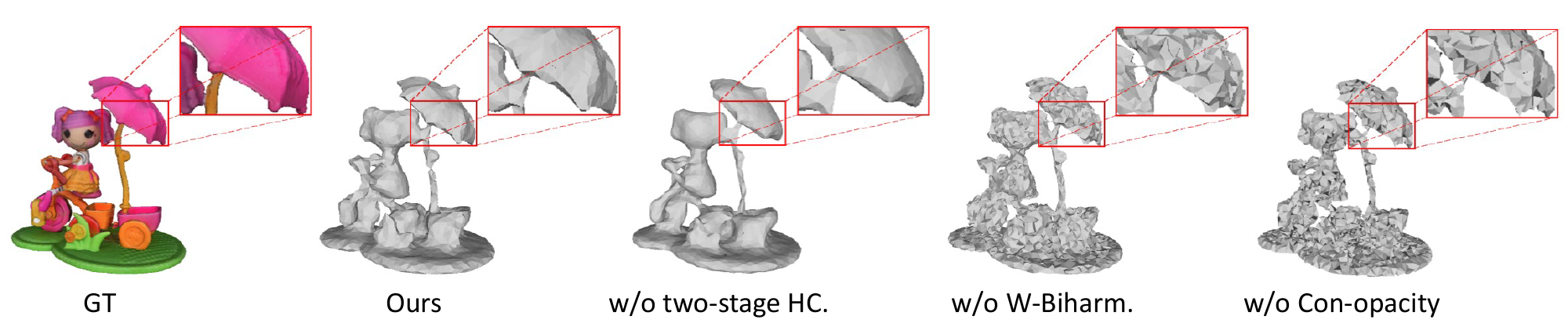}
  \vspace{-15pt}
 \caption{Qualitative ablation results. Removing the continuous opacity field leads to chaotic fragmentation, whereas disabling the two-stage HC-Laplacian or weighted bi-harmonic energy results in geometric shrinkage and the loss of high-frequency details. Our full pipeline uniquely ensures smooth, connected, and accurate reconstruction.
 }
  \label{fig:ablation}
  \vspace{-20pt}
\end{figure}

We conduct ablation studies to evaluate the individual contributions of our key design components, as summarized in \cref{table:ablation} and visualized in \cref{fig:ablation}.
First, replacing our two-stage HC–Laplacian smoothing with a conventional single-stage approach leads to pronounced surface shrinkage and over-smoothing (\cref{fig:ablation_stage}). This results in the loss of high-frequency geometric details and a corresponding decline in reconstruction accuracy.
Second, replacing our weighted bi-harmonic energy with the uniform variant reveals a geometric trade-off: although uniform regularization yields marginally higher global metrics (IoU, F1) via aggressive contraction, it cannot smooth the rough surfaces exposed after pruning, sharply degrading surface-normal continuity and edge accuracy (Normal Consistency 0.819→0.684, Edge Precision 0.379→0.178; Tab.  \ref{table:ablation}, Fig. \ref{fig:ablation}).
% Second, replacing our proposed weighted bi-harmonic energy with the original uniform bi-harmonic energy reveals a critical geometric trade-off. Although the uniform regularization yields marginally higher global proximity metrics (e.g., IoU and F1 score) due to its aggressive global contraction, it severely degrades the local surface quality. Specifically, the rough mesh surfaces exposed after topological adjustments cannot be effectively smoothed under the uniform regularization. As a result, the continuity of surface normals and the accuracy of geometric edges are heavily compromised—as evidenced by the significant drops in Normal Consistency (0.819 $\xrightarrow{}$ 0.684) and Edge Precision (0.379 $\xrightarrow{}$ 0.178). This profoundly impacts the visual and physical plausibility of the reconstructed surfaces, as clearly observed in Table \ref{table:ablation} and Figure \ref{fig:ablation}.
Finally, disabling the continuous vertex-opacity formulation makes the optimization landscape highly non-convex and shatters the mesh into 133 disconnected components, confirming its role in topological coherence.
Beyond geometry, these components also benefit rendering: since the Gaussians are bound to the mesh, removing any of them consistently lowers PSNR/SSIM and raises LPIPS (see the supplementary material).
% Finally, disabling the continuous vertex-opacity formulation results in a fragmented and topologically discontinuous mesh. Without a continuous field, the optimization landscape becomes highly non-convex, preventing effective surface refinement and inducing severe structural artifacts. Consequently, the structure shatters into numerous disconnected pieces, yielding 133 isolated tetrahedral components (see Tab.~\ref{table:ablation}).
\vspace{-10pt}
 \definecolor{myyellow}{rgb}{1,1, 0.6}
\definecolor{myorange}{rgb}{1, 0.8, 0.6}
\definecolor{myred}{rgb}{1, 0.6, 0.6}

\begin{table}[t]

\setlength{\tabcolsep}{3pt}
\begin{center}
\caption{ Quantitative ablation study on the combined 3D dataset. Disabling key components leads to distinct geometric degradation, validating their indispensable roles in our holistic framework.
}
\vspace{-10pt}
\label{table:ablation}
\small
\resizebox{\linewidth}{!}{
\begin{tabular}{lcccc}
\toprule
   Metrics  &  {w/o two-stage HC.} &
 \multicolumn{1}{c}{w/o W-Biharm.}&
 \multicolumn{1}{c}{w/o Con-opacity}&
 \multicolumn{1}{c}{{\textbf{Ours}}}\\
\midrule
Normal Consistency  $ \uparrow$  & 0.784 &0.684 & 0.597 & \textbf{0.819 } \\
F1 score  $ \uparrow$  & 0.317 & \textbf{0.444} & 0.405 & 0.428  \\
Edge F1 score $\uparrow$ & 0.132 & 0.274 & 0.202&  \textbf{0.285}  \\ 
Edge precision $\uparrow$ & 0.227 & 0.178 & 0.119&  \textbf{0.379}  \\ 
IOU  $\uparrow$ & 0.558 & \textbf{0.645} & 0.585&  0.620 \\ 
Max Connection Rate $\uparrow$ & \textbf{1.0} & \textbf{1.0 }& 0.98&  \textbf{1.0 } \\ 
Num. components $\downarrow$ &  \textbf{1} &  \textbf{1} & 133 &   \textbf{1} \\

\bottomrule
\end{tabular}
}
\end{center}

\vspace{-25pt}
\end{table}

\section{Conclusions}
In this paper, we introduce HoloTetSphere, a topology-adaptive and holistic volumetric representation. Building upon the novel Lagrangian framework of TetSphere, our approach effectively overcomes its critical reliance on strict initializations and the structural limitations inherent to discrete primitives. 
By integrating topological and geometric optimization, HoloTetSphere produces unified and topologically coherent tetrahedral meshes of high geometric fidelity that, after a lightweight automatic element cleanup, provide a practical geometric foundation for surface-level physical simulation of uniform solid media.
%By seamlessly integrating topological and geometric optimization, HoloTetSphere produces unified and topologically coherent tetrahedral meshes of high geometric fidelity, providing a highly compatible geometric foundation for realistic physical simulations.

\noindent\textbf{Limitations.} 
Under strict orientation-preserving regularization, highly complex topologies can occasionally cause localized element inversion; as discussed in Sec. \ref{sec:physim}, a lightweight automatic cleanup resolves these, though extreme cases may still need in-simulator handling. Moreover, our pruning-based adaptation removes rather than adds material, targeting single-object connected reconstruction rather than arbitrary topology editing. Very thin, plate-like structures also remain challenging: reconstruction stays accurate down to a relative thickness of about $0.34$ and degrades below this threshold (see the supplementary material).
% Since HoloTetSphere optimizes continuous tetrahedral meshes under strict orientation-preserving regularization, adapting to highly complex topologies can occasionally demand extreme elemental deformations. This may lead to localized element inversion. Consequently, minor restorative post-processing within the physical simulator (e.g., inverted element handling) may be required to ensure unconditionally stable simulation dynamics.

% ---------------------------------------------------------------
% Acknowledgements (camera-ready: place on page 15 per ECCV instructions)
\section*{Acknowledgements}
This work is supported in part by the NSFC (62132021), the Major Program of Xiangjiang Laboratory (23XJ01009), NSFC (62572477, 62522219, 62372457, 62325211), and the Research Fund of Jiangsu Key Laboratory of AI for Industries (E6420016G8).

% ---------------------------------------------------------------
% Bibliography
\bibliographystyle{splncs04}
\bibliography{main}

\clearpage
\appendix

% =====================================================================
\section{Overview}
\label{sec:supp_overview}
% =====================================================================

This supplementary material provides additional implementation details,
extended experiments, and physical simulation configurations that complement
the main paper.  The content is organized as follows:
Section~\ref{sec:implementation} gives comprehensive hyperparameter settings
to ensure full reproducibility;
Section~\ref{sec:simulation_details} outlines the physical simulation
setup;
Section~\ref{sec:additional_results} reports extended experimental results
for both rendering and geometry comparisons, and presents additional qualitative results. 
Section~\ref{sec:limitations} discusses limitations, failure cases, and
future directions.

% =====================================================================
\section{Implementation Details}
\label{sec:implementation}
% =====================================================================

We provide comprehensive implementation details of the \emph{HoloTetSphere}
framework, which consists of three main stages: initialization,
differentiable topology optimization, and coupled geometry optimization.

% ------------------------------------------------------------------
\subsection{Initialization}
% ------------------------------------------------------------------

\subsubsection{Initial Gaussian Representation.}
We do not rely on external Structure-from-Motion (SfM) points for
initialization.  Instead, we first establish a coarse 3D representation
using rapid 2D Gaussian Splatting~\cite{huang20242d} to capture the
overall object structure.  Detailed training settings are as follows.

\noindent\textbf{Training duration}: 10{,}000 iterations.

\noindent\textbf{Loss function}: Photometric reconstruction loss:
\begin{equation}
\mathcal{L}_{\text{init}}
  = (1-\lambda_{\text{DSSIM}})\,\mathcal{L}_1
  + \lambda_{\text{DSSIM}}\,(1-\text{SSIM}),
\end{equation}
where $\lambda_{\text{DSSIM}} = 0.2$ by default.

\noindent\textbf{Densification}: Follows the standard 2DGS protocol with
adaptive density control and opacity reset at specified intervals.

\subsubsection{Convex Hull-based Tetrahedral Initialization.}
After obtaining the initial Gaussian point cloud, we construct the
tetrahedral mesh initialization through the following steps.

\noindent\textbf{Point cloud extraction}: Extract Gaussian centers
$\{\mathbf{g}_i\}_{i=1}^{N}$ from the trained model.

\noindent\textbf{Convex hull computation}: Compute the 3-D convex hull
$\mathcal{H}$ from the Gaussian centers using Open3D.

\noindent\textbf{Volumetric scaling}: Scale $\mathcal{H}$ by a factor of
$1.1$ from its centroid to guarantee strict bounding coverage of the target
object.

\noindent\textbf{Tetrahedralization}: Generate a coarse tetrahedral mesh
$\mathcal{M} = \{V_0, T_0\}$ from $\mathcal{H}$ using TetWild (with
$\epsilon=0.001$), where $V_0 = \{\mathbf{v}_i\}_{i=1}^{N_v}$ are vertices
and $T_0 = \{t_j\}_{j=1}^{N_t}$ are tetrahedral elements.

\noindent\textbf{Gaussian--Tetrahedra coupling}: Initialize  \emph{one}
Gaussian sphere per tetrahedral element, strictly positioned at the
element's centroid (1:1 correspondence, no further densification is
performed in this stage).

This initialization strategy endows the resulting mesh with three desirable
properties: (i) it starts as a single, globally connected structure;
(ii) it provides full spatial coverage of the target object; and
(iii) it offers a smooth transition from a point-based representation
to a volumetric one.

% ------------------------------------------------------------------
\subsection{Differentiable Topology Optimization}
% ------------------------------------------------------------------

\subsubsection{Relationship between the Conceptual Objective and the
Actual Implementation.}
The main paper (Eq.~2) introduces a high-level objective
$\mathcal{L}_{\text{conceptual}} = \mathcal{D}_{\text{topo}}(\hat{T},T) +
\mathcal{D}_{\text{geom}}(\hat{G},G)$ as an organizing principle.
In practice, the topology term is realized by the continuous opacity field
trained with $\mathcal{L}_{\text{topo}}$ (Eq.~5 of the main paper), and the
geometry term is realized by the joint geometry loss
$\mathcal{L}_{\text{geo}}$ together with all regularizer (Eq.~9 of the
main paper).
The two sub-objectives are executed in an alternating schedule rather than
simultaneously, because joint differentiation through both the tetrahedral
topology removal and the Gaussian rasteriser in a single backward pass
may cause numerical instabilities.

\subsubsection{Topology Optimization Loss.}
The topology optimization objective follows Eq.~5 of the main paper:
\begin{equation}
\mathcal{L}_{\text{topo}}
  = \mathcal{L}_{\text{render}}
  + \lambda_{\text{eik}}\,\mathcal{L}_{\text{eik}}
  + \lambda_{\text{smooth}}\,\mathcal{L}_{\text{smooth}},
\end{equation}
where the regularization terms are defined over the mesh edge set
$\mathcal{E}$ as (Eq.~6 of the main paper):
\begin{equation}
\mathcal{L}_{\text{eik}}
  = \frac{1}{|\mathcal{E}|}
    \sum_{(u,v)\in\mathcal{E}}
    \left(\frac{|\phi_u - \phi_v|}{\ell_{uv}} - 1\right)^{\!2},
\qquad
\mathcal{L}_{\text{smooth}}
  = \frac{1}{|\mathcal{E}|}
    \sum_{(u,v)\in\mathcal{E}}
    (\phi_u - \phi_v)^{2},
\end{equation}
The loss weights are empirically set as
$\lambda_{\text{eik}} =3.0$, $\lambda_{\text{smooth}} = 3.0$.

During the \textbf{Gaussian rendering phase} (iterations 0--15{,}000),
the following auxiliary losses supervise the Gaussian spheres coupled to
tetrahedral elements.  These losses operate \emph{exclusively} on Gaussian
attributes and are \emph{not} part of $\mathcal{L}_{\text{geo}}$
(Eq.~9 of the main paper).  The total loss is assembled as:
\begin{equation}
\mathcal{L}_{\text{render}}
  = \lambda_{\text{rgb}}\,\mathcal{L}_{\text{rgb}}
  + \lambda_{\text{mask}}\,\mathcal{L}_{\text{mask}}
  + \lambda_{\text{normal}}\,\mathcal{L}_{\text{normal}}
  + \lambda_{\text{prior}}\,\mathcal{L}_{\text{prior}},
\end{equation}
where $\lambda_{\text{rgb}} = 10.0$ , $\lambda_{\text{mask}} = 3.0$ , $\lambda_{\text{normal}} = 0.05$ and $\lambda_{\text{prior}} = 0.25$.

\noindent\textbf{RGB Loss}:
\begin{equation}
\mathcal{L}_{\text{rgb}}
  = (1-\lambda_{\text{DSSIM}})\,\mathcal{L}_1
  + \lambda_{\text{DSSIM}}\,(1-\text{SSIM}).
\end{equation}

\noindent\textbf{Mask Loss} (corresponding to $\mathcal{L}_{\text{mask}}$
in Eq.~9 of the main paper):
\begin{equation}
\mathcal{L}_{\text{mask}}
  = \|\hat{\mathcal{O}} - M_{\text{gt}}\|_1,
\end{equation}
where $\hat{\mathcal{O}}$ is the rendered opacity map and $M_{\text{gt}}$
is the foreground silhouette mask.

\noindent\textbf{Normal Consistency Loss}:
\begin{equation}
\mathcal{L}_{\text{normal}}
  = \mathcal{R}\!\left(
    \mathcal{E}(\mathbf{n}^{\text{GS}}, \mathbf{n}^{\text{rend}})
    \right),
\end{equation}
where $\mathcal{E}(\mathbf{n}_1, \mathbf{n}_2) = 1 - \cos(\mathbf{n}_1,
\mathbf{n}_2)$, and $\mathcal{R}(\cdot)$ denotes the
ranking loss that selectively penalizes large errors.

\noindent\textbf{Normal Prior Loss}:
\begin{equation}
\mathcal{L}_{\text{prior}}
  = \mathcal{R}\!\left(
    \mathcal{E}(\mathbf{n}_{\text{prior}}, \mathbf{n}^{\text{GS}})
    + \mathcal{E}(\mathbf{n}_{\text{prior}}, \mathbf{n}^{\text{rend}})
    \right).
\end{equation}
Here $\mathbf{n}_{\text{prior}}$ denotes a per-pixel surface normal
obtained by running a monocular normal estimator on each input view during
preprocessing; the resulting normal maps are stored alongside the camera
data and loaded at training time.

\subsubsection{Adaptive Topology Pruning.}
\label{sec:pruning}
\noindent\textbf{Opacity computation.}
For each tetrahedral element $t_i$ with vertex scalar field values
$\{\phi_{i,j}\}_{j=1}^{4}$, the element-level opacity follows
Section~3.3 of the main paper:
\begin{equation}
\bar{\phi}_i = \frac{1}{4}\sum_{j=1}^{4} \phi_{i,j},
\qquad
\alpha_i = \sigma\!\left(-s\cdot\bar{\phi}_i\cdot\kappa\right),
\end{equation}
where $s$ is a \emph{fixed} unit-scale factor ($s = 1.0$), $\kappa$ is a
\emph{learnable} sharpness parameter that controls the transition steepness
between inside and outside regions, and $\sigma(\cdot)$ is the sigmoid
function.
$\kappa$ is initialized to $10.0$ and optimized with a dedicated Adam
instance at learning rate $0.01$.
Intuitively, large $\kappa$ sharpens the pruning boundary, whereas small
$\kappa$ yields a smooth, differentiable transition.
To prevent training instability, the effective product $\kappa' =
s\cdot\kappa$ is clamped to the range $[1.0,\,50.0]$ throughout
optimization.

\noindent\textbf{Pruning schedule.}
To maintain stable gradient flow, topology pruning is performed at iteration 15{,}000, 20{,}000, and 25{,}000, when optimization has converged
sufficiently for opacity values to be reliable.
Elements whose average opacity satisfies $\alpha_i < \tau_\alpha = 0.1$
are identified as redundant and permanently removed, updating
$\Omega_{\text{valid}}$ as defined in Eq.~4 of the main paper.
After pruning, the Gaussian--tetrahedra correspondence is updated and the
representation is re-initialized as follows.
\textbf{Color and rotation} attributes of surviving Gaussians are
\emph{inherited} directly from their pre-pruning values via the tet-level
validity mask.
\textbf{Scale} is \emph{recomputed} from the updated tetrahedral geometry:
the scale factor for each Gaussian is set to
$s_k = 0.65 \cdot \bar{d}_k$,
where $\bar{d}_k$ is the mean distance from the host tetrahedron's centroid
to its four vertices; the factor $0.65$ is a fixed empirical constant.
After this recomputation, the scale parameter is converted from a fixed
quantity to a \emph{learnable} parameter (Adam, lr\,$= 0.005$) to allow
further refinement.
\textbf{Opacity} is converted from the SDF-derived value to a
\emph{learnable} parameter (Adam, lr\,$= 0.05$), initialized from the
current sigmoid mapping of the vertex SDF field, enabling the network to
freely adjust opacity during subsequent geometry optimization.

% ------------------------------------------------------------------
\subsection{Coupled Geometry Optimization}
% ------------------------------------------------------------------

The full pipeline consists of three sequential stages.
\textbf{Stage 0} (pre-training): a standalone 2DGS model is trained for
10{,}000 iterations to obtain the Gaussian point cloud used for convex-hull
initialization.
\textbf{Stage 1} (main loop): 30{,}000 Gaussian rendering iterations,
during which topology pruning is triggered at iteration 15{,}000, 20{,}000 and 25{,}000.
\textbf{Stage 2} (mesh optimization blocks): at iterations 20{,}000 and
25{,}000, Gaussian rendering attributes are frozen and 5{,}000 iterations
of pure tetrahedral vertex optimization are inserted; these two blocks
contribute approximately 10{,}000 mesh-optimization steps in total.
The complete training schedule is summarized in
Table~\ref{tab:time_breakdown}.

\subsubsection{Tetrahedral Mesh Optimization.}
The joint geometry optimization follows Eq.~9 of the main paper:
\begin{equation}
\mathcal{L}_{\text{geo}}
  = \lambda_m\,\mathcal{L}_{\text{mask}}
  + \lambda_n\,\mathcal{L}_{\text{norm}}
  + \lambda_{\text{HC}}\,\mathcal{L}_{\text{HC}}
  + \lambda_w\,\mathcal{L}_{w},
\end{equation}
with $\lambda_m = 20$ and $\lambda_n = 0.25$.
The regularization weights $\lambda_{\text{HC}}$ and $\lambda_w$ follow a
decaying schedule over the mesh-optimization phase:
\begin{equation}
\lambda_{\text{HC}} = \lambda_w
  = \left(1 - \frac{t}{10{,}000}\right)^{\!2},
\end{equation}
where $t$ denotes the current mesh-optimization iteration.
The base inner coefficients are
$c_{\text{smooth}} = 2\times10^{-4}$ and $c_{\text{barrier}} = 2\times10^{-4}$,
each normalized by the number of tetrahedral spheres in the scene.
This decay schedule reduces regularization pressure as the mesh converges,
allowing finer geometric detail to emerge in later iterations.

\noindent\textbf{Weighted Bi-harmonic Regularization}
($\mathcal{L}_w$ in Eq.~9 of the main paper):
\begin{equation}
\mathcal{L}_{w}
  = w(\mathbf{x})\,\lVert \mathbf{L}\mathbf{F}_{\mathbf{x}} \rVert_2^2,
\end{equation}
with the weight function
\begin{equation}
w(\mathbf{v}_i)
= \begin{cases}
5\times 10^{-4}, & \mathbf{v}_i \in \mathcal{S}
                  \text{ (surface vertices)}\\
1.0,             & \mathbf{v}_i \in \mathcal{V}\setminus\mathcal{S}
                  \text{ (interior vertices),}
\end{cases}
\end{equation}
where $\mathcal{S}$ is the surface vertex set defined in Eq.~8 of the main
paper ($\gamma = 5\times10^{-4} \ll 1$).
Surface vertices $\mathcal{S}$ are \emph{dynamically} identified at each
mesh-optimization step by extracting boundary triangular faces,
i.e.\ faces that appear exactly once in the current tetrahedral
connectivity.
This weighted formulation gives surface vertices greater freedom of
deformation to fit fine geometric details, while maintaining strong
structural regularization for interior vertices.

\noindent\textbf{Two-stage HC-Laplacian Smoothing}
($\mathcal{L}_{\text{HC}}$ in Eq.~9 of the main paper):
We employ the filtered HC-Laplacian energy defined over surface vertices
$\mathcal{S}$ following Eq.~7 of the main paper, decomposed into
tangential ($d^t$) and normal ($d^n$) displacement components:
\begin{equation}
\mathcal{L}_{\text{HC}}
  = \sum_{i\in\mathcal{S}}
    \left[
      \|d_i^{t}\|^{2}
      + \lambda_{\text{cond}}(t)\,(d_i^{n})_{-}^{2}
      + (d_i^{n})_{+}^{2}
    \right],
\end{equation}
where $(d_i^n)_-$ and $(d_i^n)_+$ denote the inward and outward normal
displacement components respectively, and the stage transition occurs at
$t=5{,}000$ steps of the mesh-optimization phase:
\begin{equation}
\lambda_{\text{cond}}(t)
= \begin{cases}
0.0, & \text{Stage~1}\;(t < 5000){:}
      \text{ push out concave regions only}\\
1.0, & \text{Stage~2}\;(t \ge 5000){:}
      \text{ overall global smoothing.}
\end{cases}
\end{equation}
In Stage~1, inward-pointing normal displacements are not penalized,
allowing concave regions to be pushed outward and corrected.
In Stage~2, the full HC energy is active, providing global smoothing
without the mesh-shrinkage artifact of standard Laplacian smoothing.

\noindent\textbf{Gaussian-guided Normal Supervision}
($\mathcal{L}_{\text{norm}}$ in Eq.~9 of the main paper):
Using the (frozen) Gaussian spheres as surface-normal supervision,
following Section~3.4 of the main paper:
\begin{equation}
\mathcal{L}_{\text{norm}}
  = \sum_{p} M_{\text{mask}}(p)
    \left(1 - \langle \mathbf{n}^{\text{GS}}_p,\,\mathbf{n}^{\text{Tet}}_p \rangle\right),
\end{equation}
where $M_{\text{mask}}(p)$ is a binary mask selecting pixels with
non-zero rendered alpha, restricting normal supervision to visible surface
regions, and $\mathbf{n}^{\text{Tet}}$ denotes the rasterized normals of
the tetrahedral mesh.

\subsubsection{Tetrahedral--Gaussian Update Coupling.}
After each tetrahedral optimization phase, each Gaussian sphere's position
is updated via barycentric interpolation from its host tetrahedron's
(possibly moved) vertices, following Section~3.4 of the main paper:
\begin{equation}
\mathbf{g}_k^{\text{new}}
  = \sum_{i=1}^{4} \beta_{ik}\,\mathbf{v}_i^{\text{new}},
  \qquad
  \beta_{ik} \ge 0,\quad \sum_{i=1}^{4}\beta_{ik} = 1,
\end{equation}
where $\beta_{ik}$ are the barycentric coordinates of the $k$-th Gaussian
center with respect to the vertices of its host tetrahedron, computed once
at coupling time and held fixed throughout each 5{,}000-step
mesh-optimization block.
Gaussian rendering attributes (color, opacity, scale, rotation) are
\emph{not} updated during the tetrahedral phase; only positions change.
Note that this applies within each mesh-optimization block.
The one-time post-pruning re-initialization of scale and opacity
described in Section~\ref{sec:pruning} is a separate procedure that
occurs only at iteration 15{,}000 and does not repeat during subsequent
mesh-optimization phases.

% ------------------------------------------------------------------
\subsection{Optimizers and Computational Cost}
% ------------------------------------------------------------------

We use the Adam optimizer for Gaussian attributes and an independent
uniform Adam optimizer for tetrahedral vertices.
Gaussian positions employ exponential learning-rate decay from
$1.6\times10^{-5}$ to $1.6\times10^{-6}$.
During the 5{,}000-step geometric optimization phases, the tetrahedral
vertex learning rate is initialized at $2\times10^{-4}$ and decays
following a Cosine-Annealing schedule to $1\times10^{-4}$.

All experiments are conducted on a single NVIDIA RTX 4090 GPU (24\,GB
VRAM).  A typical scene requires approximately \textbf{40~minutes} to
complete the full 30{,}000-iteration pipeline: roughly 5~minutes for the
initial 2DGS stage, 15~minutes for the differentiable topology optimization
stage, and 23~minutes for the two alternating
mesh-optimization phases.
Table~\ref{tab:time_breakdown} summarizes the per-stage timing.

\begin{table}[t]
\setlength{\tabcolsep}{6pt}
\centering
\small
\caption{Per-stage training time on a single NVIDIA RTX 4090 (typical
         scene, single object). The 2DGS initialization is a pre-training
         stage; the remaining 30{,}000 iterations form the main pipeline.}
\label{tab:time_breakdown}
\begin{tabular}{lccc}
\toprule
Stage & Iterations & Time (min) & VRAM (GB) \\
\midrule
2DGS initialization              & 10{,}000 & $\approx$5  & 3.2 \\
Topology optimization            & 15{,}000 & $\approx$15 & 6.1 \\
Gaussian-guided optimization     & 15{,}000 & $\approx$11  & 6.1 \\
Mesh-guided optimization    & 10{,}000 & $\approx$12 & 7.6 \\
\bottomrule
\end{tabular}
\end{table}

\definecolor{myyellow}{rgb}{1,1, 0.6}
\definecolor{myorange}{rgb}{1, 0.8, 0.6}
\definecolor{myred}{rgb}{1, 0.6, 0.6}

\setlength{\tabcolsep}{4pt}
\begin{table}[t]
\caption{
Quantitative PSNR comparison between our method and competing baselines on the DTU dataset. Our method attains the highest rendering quality across all scenes.}
\label{table:dtu}
\renewcommand\arraystretch{1.25}
\begin{center}
\resizebox{\linewidth}{!}{
\begin{tabular}{c|cccccc}

\hline
 ScanID & 
 Voxurf  & Neus & Instant-NGP & Instant-NSR &
 Neus2  & \textbf{ Ours}
 \\
\hline
scan24 & 27.89 &  26.49  & 28.32 & 23.86&  {28.44}  & \textbf{ 31.95} \\     
scan37 & 26.90 & 26.17  & 27.19  & 24.97 & 26.53 & \textbf{29.08} \\   
scan40 & 28.81 &  27.66 &30.45& 25.3 & {29.70}  & \textbf{31.78} \\   
scan55 & 31.02 &  27.78  &29.81& 25.43& {31.47} &  \textbf{33.94} \\   
scan63 & {34.38} &  30.63 &31.22& 29.52 & 33.74 &  \textbf{37.69} \\   
scan65 & {31.48} &  27.42 &27.78&26.17& 30.99 & \textbf{35.05}\\   
scan69 & {30.13} &  25.83 &24.79& 22.93 & 28.77&\textbf{34.87} \\   
scan83 & {37.43} &  30.00 &31.23& 26.72& 36.78 &  \textbf{39.90} \\   
scan97 & {28.35} & 26.40 &26.96 &25.94& 28.24 &  \textbf{32.52} \\   
scan105 & 32.94 &  29.58 &30.62&27.71& 29.63  & \textbf{36.38}\\   
scan106 & {34.17} &  25.87 &25.62&23.12& {33.91} &  \textbf{37.61}\\   
scan110 & 32.70 &  28.82 &28.6&25.44& {34.50} & \textbf{36.22} \\   
scan114 & 30.97 &  28.80 &29.5&26.7& {31.14}  & \textbf{32.98} \\   
scan118 & {37.24} &  27.36 &27.91&25.13& {37.17} & \textbf{40.31} \\   
scan122 & {37.97} & 31.19 &32.93&28.19& {37.41} & \textbf{40.18}\\   
\hline
mean &  {32.16} &  28.00 &28.86& 25.81&32.14 & \textbf{ 35.36} \\ 
\hline
\end{tabular}
}
\end{center}
\end{table}

% =====================================================================
\section{Physical Simulation Details}
\label{sec:simulation_details}
% =====================================================================

\subsection{Simulation Framework}
We conduct all physical simulations using NVIDIA Omniverse Isaac Sim with
PhysX~5.0 GPU-accelerated deformable body simulation.
Our implementation uses tetrahedral Finite Element Method (FEM) with
position-based dynamics (PBD) for stable real-time performance.

\subsection{Numerical Integration Settings}
\noindent\textbf{Time step}: $\Delta t = 1/60$\,s (60 FPS).

\noindent\textbf{Solver iterations}: 16 position-based iterations per step.

\noindent\textbf{Collision detection}: Continuous Collision Detection (CCD)
enabled.

\noindent\textbf{Self-collision}: Disabled to improve computational
efficiency.

\noindent\textbf{GPU acceleration}: PhysX GPU dynamics with TGS solver.

\noindent\textbf{Broadphase}: GPU-based spatial partitioning.

\subsection{Scene Setup}
\noindent\textbf{Coordinate system}: Z-up with configurable scene rotation.

\noindent\textbf{Lighting}: Distant (sun) light at 45$^\circ$ elevation
plus a dome light for ambient illumination.

\noindent\textbf{Rendering}: 1280$\times$720 viewport capture at 60~FPS.

\noindent\textbf{Output}: Sequential frame capture for qualitative analysis.

% =====================================================================
\section{Additional Experimental Results}
\label{sec:additional_results}
% =====================================================================

\subsection{Additional Qualitative Results}
\begin{figure}[t]
  \centering
  \includegraphics[width=\linewidth]{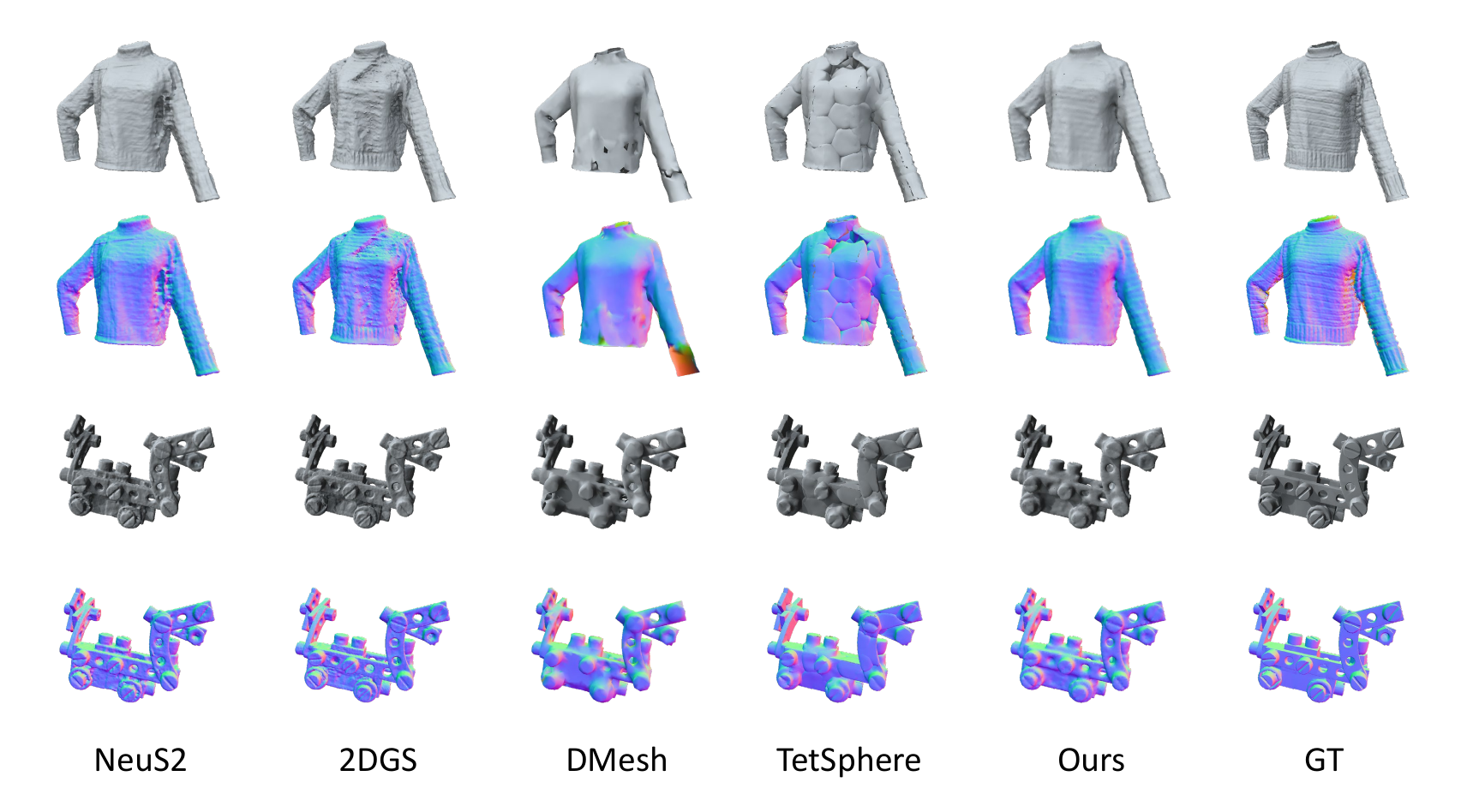}
  \vspace{-6px}
  \caption{Qualitative comparisons of reconstructed meshes and normal maps
           (additional scenes, part 1).}
  \label{fig:geo1}
  \vspace{-6px}
\end{figure}

\begin{figure}[t]
  \centering
  \includegraphics[width=\linewidth]{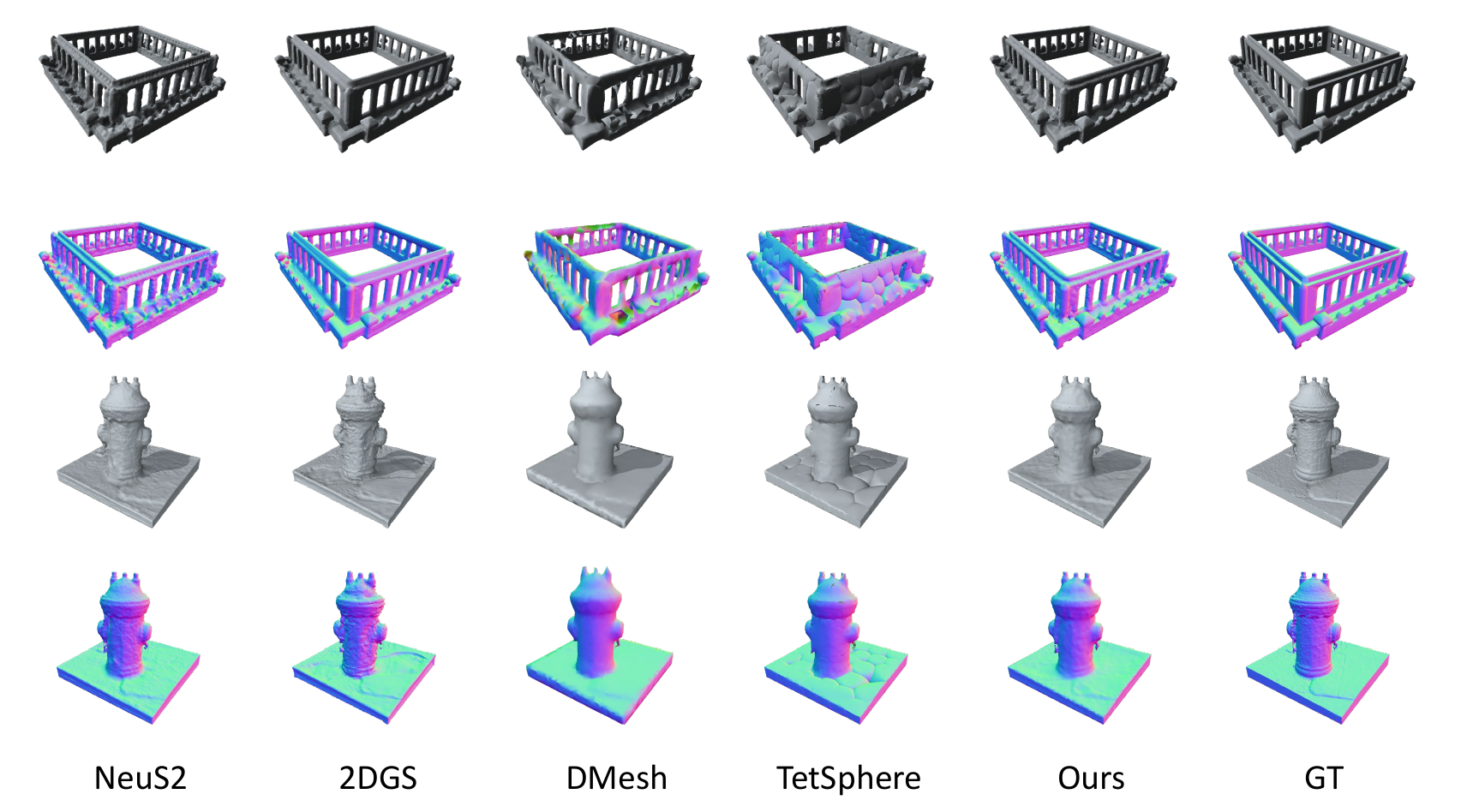}
  \vspace{-6px}
  \caption{Qualitative comparisons of reconstructed meshes and normal maps
           (additional scenes, part 2).}
  \label{fig:geo2}
  \vspace{-6px}
\end{figure}

\begin{figure}[t]
  \centering
  \includegraphics[width=\linewidth]{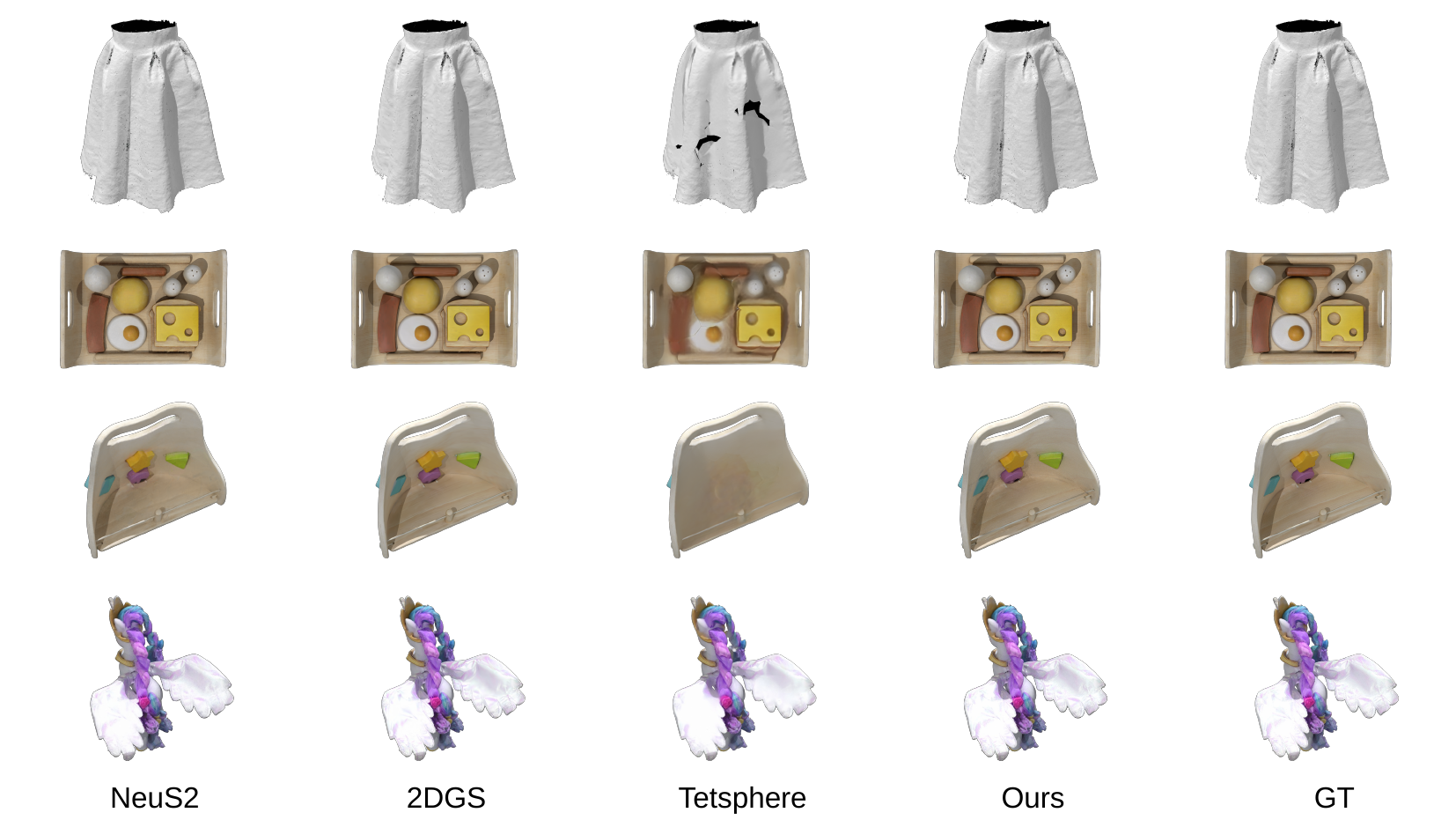}
  \vspace{-6px}
  \caption{Novel-view rendering comparisons (part 1).}
  \label{fig:rend1}
  \vspace{-6px}
\end{figure}

\begin{figure}[t]
  \centering
  \includegraphics[width=\linewidth]{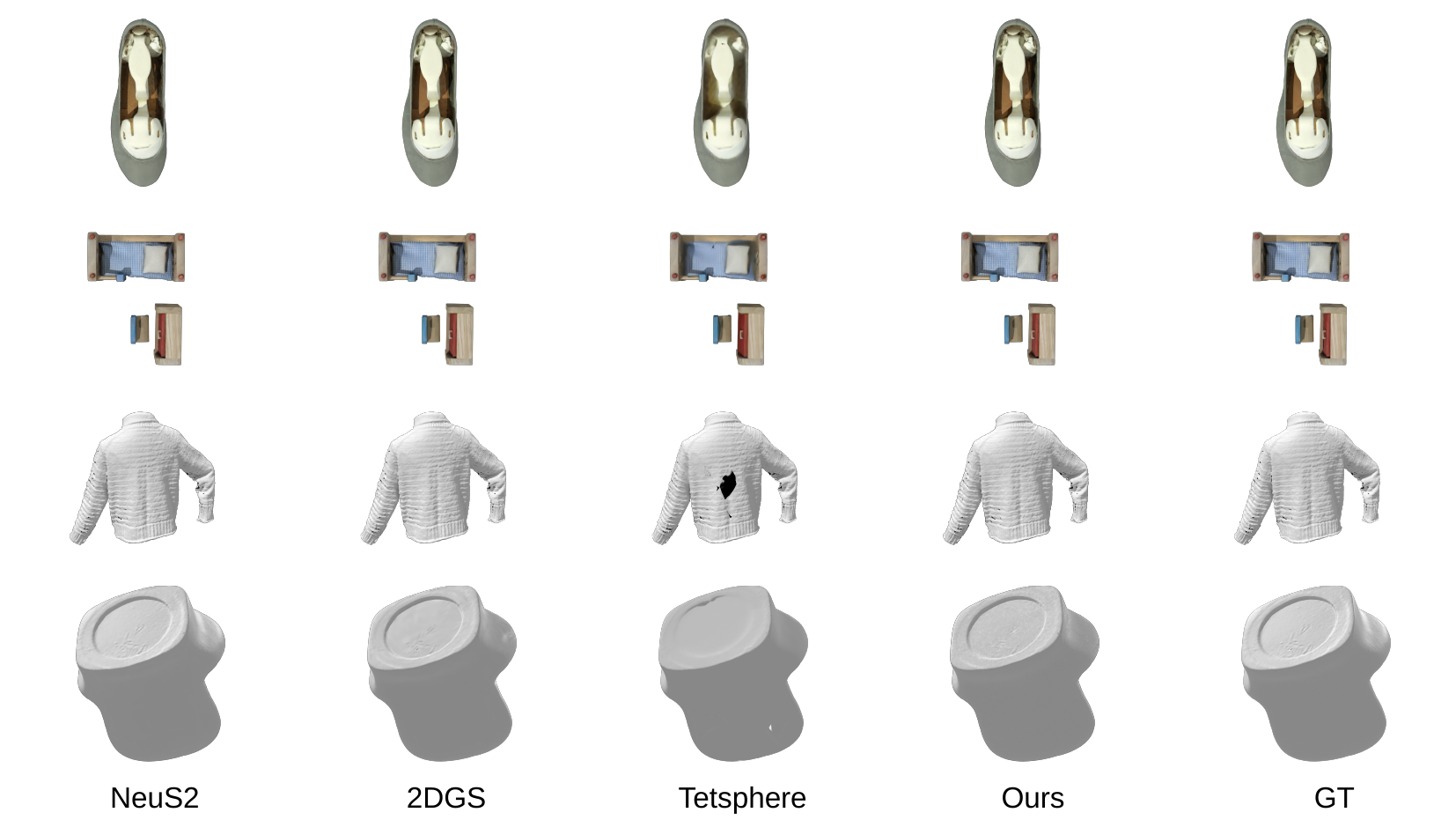}
  \vspace{-6px}
  \caption{Novel-view rendering comparisons (part 2).}
  \label{fig:rend2}
  \vspace{-6px}
\end{figure}

In Fig.~\ref{fig:geo1} and Fig.~\ref{fig:geo2} we provide more qualitative
results of reconstructed meshes and normal maps, comparing NeuS2~\cite{neus2},
2DGS~\cite{huang20242d}, DMesh~\cite{son2024dmesh},
TetSphere Splatting~\cite{guo2024tetsphere}, and our method.
In Fig.~\ref{fig:rend1} and Fig.~\ref{fig:rend2} we provide more
qualitative results of novel-view rendering, comparing
NeuS2~\cite{neus2}, 2DGS, TetSphere Splatting, and our method.


\subsection{Additional Quantitative Results}
For a more comprehensive assessment of rendering quality, we further include the DTU dataset~\cite{jensen2014large} with 15 multi-view scenes, each containing either 49 or 64 images.
For view-synthesis evaluation on the DTU dataset, we compare against voxel- or grid-based neural rendering methods, including Voxurf~\cite{wu2022voxurf}, Instant-NGP~\cite{muller2022instant}, and Instant-NSR~\cite{zhao2022human}.

Benefiting from the joint optimization of tetrahedra and Gaussians, our method inherits the strong rendering capability of Gaussian splatting, leading to significantly higher view-synthesis quality compared with implicit neural rendering approaches, as shown in Tab. \ref{table:dtu}.  
In addition, the geometric consistency enforced by our tetrahedral mesh—particularly the uniformity of tetrahedral elements—provides more stable and accurate Gaussian placements.

\paragraph{Rendering ablation.}
Because the Gaussians are barycentrically coupled to the tetrahedral mesh, the topology-aware optimization and regularization that improve geometric quality concurrently enhance rendering fidelity. To verify this, we additionally re-run the same ablation variants used in the main paper and report their rendering metrics in \cref{tab:render_ablation}. Removing any of these terms consistently degrades both PSNR/SSIM and LPIPS, confirming this joint benefit; the weighted bi-harmonic energy has a comparatively smaller rendering impact because it mainly regularizes near-surface tetrahedra.

\begin{table}[t]
\centering
\scriptsize
\caption{Ablation study with rendering metrics, using the same variants as the geometry ablation in the main paper.}
\label{tab:render_ablation}
\begin{tabular*}{\linewidth}{@{\extracolsep{\fill}}lccc}
\toprule
\textbf{Variant} & \textbf{PSNR} $\uparrow$ & \textbf{SSIM} $\uparrow$ & \textbf{LPIPS} $\downarrow$ \\
\midrule
Ours & \textbf{40.9509} & \textbf{0.99036} & \textbf{0.01896} \\
w/o W-Biharm. & 40.3958 & 0.98909 & 0.02123 \\
w/o two-stage HC. & 39.3252 & 0.98773 & 0.02323 \\
w/o Con-opacity & 39.5411 & 0.98685 & 0.02675 \\
\bottomrule
\end{tabular*}
\end{table}

\subsection{Additional Physical Simulation Results}
\begin{figure}[t]
  \centering
  \includegraphics[width=\linewidth]{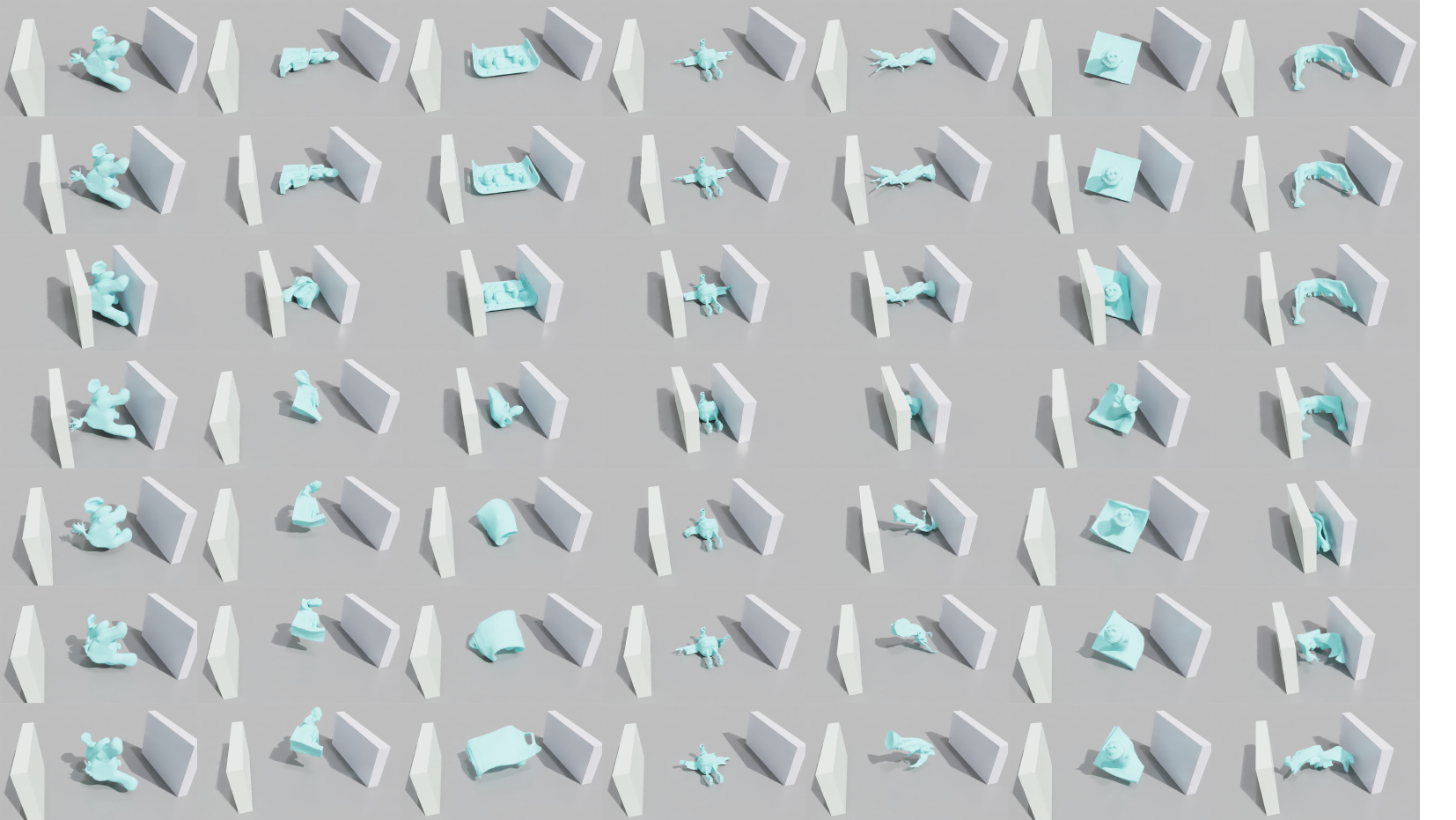}
  \vspace{-6px}
  \caption{Additional physical simulation results (part 1).
           Our single-component tetrahedral mesh enables stable
           deformable-body simulation throughout the sequence.}
  \label{fig:phy1}
  \vspace{-6px}
\end{figure}

\begin{figure}[t]
  \centering
  \includegraphics[width=\linewidth]{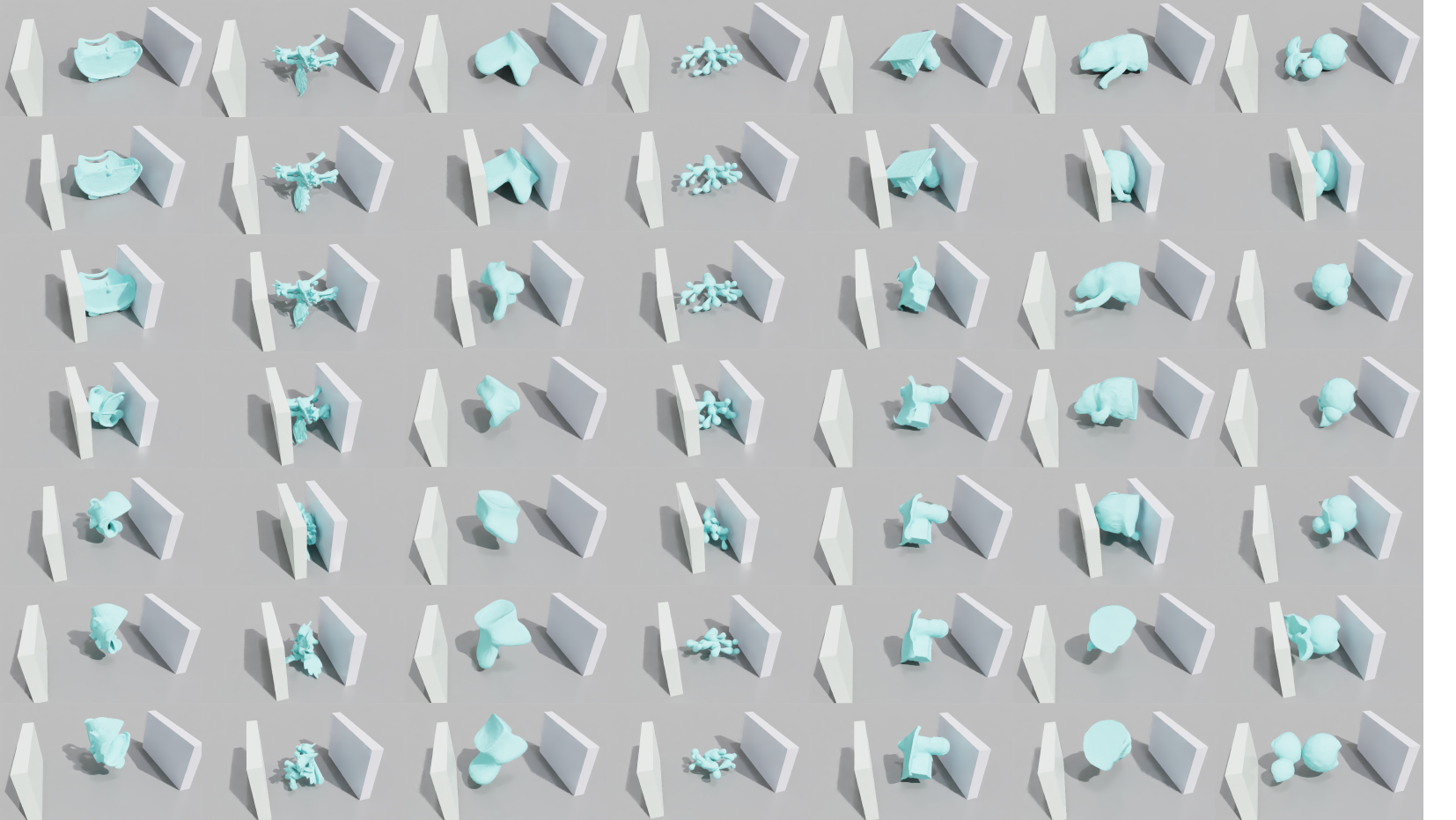}
  \vspace{-6px}
  \caption{Additional physical simulation results (part 2).}
  \label{fig:phy2}
  \vspace{-6px}
\end{figure}

We visualize additional reconstruction results on the simulation platform
in Fig.~\ref{fig:phy1} and Fig.~\ref{fig:phy2}.

\begin{figure}
  \centering
  \includegraphics[width=\linewidth]{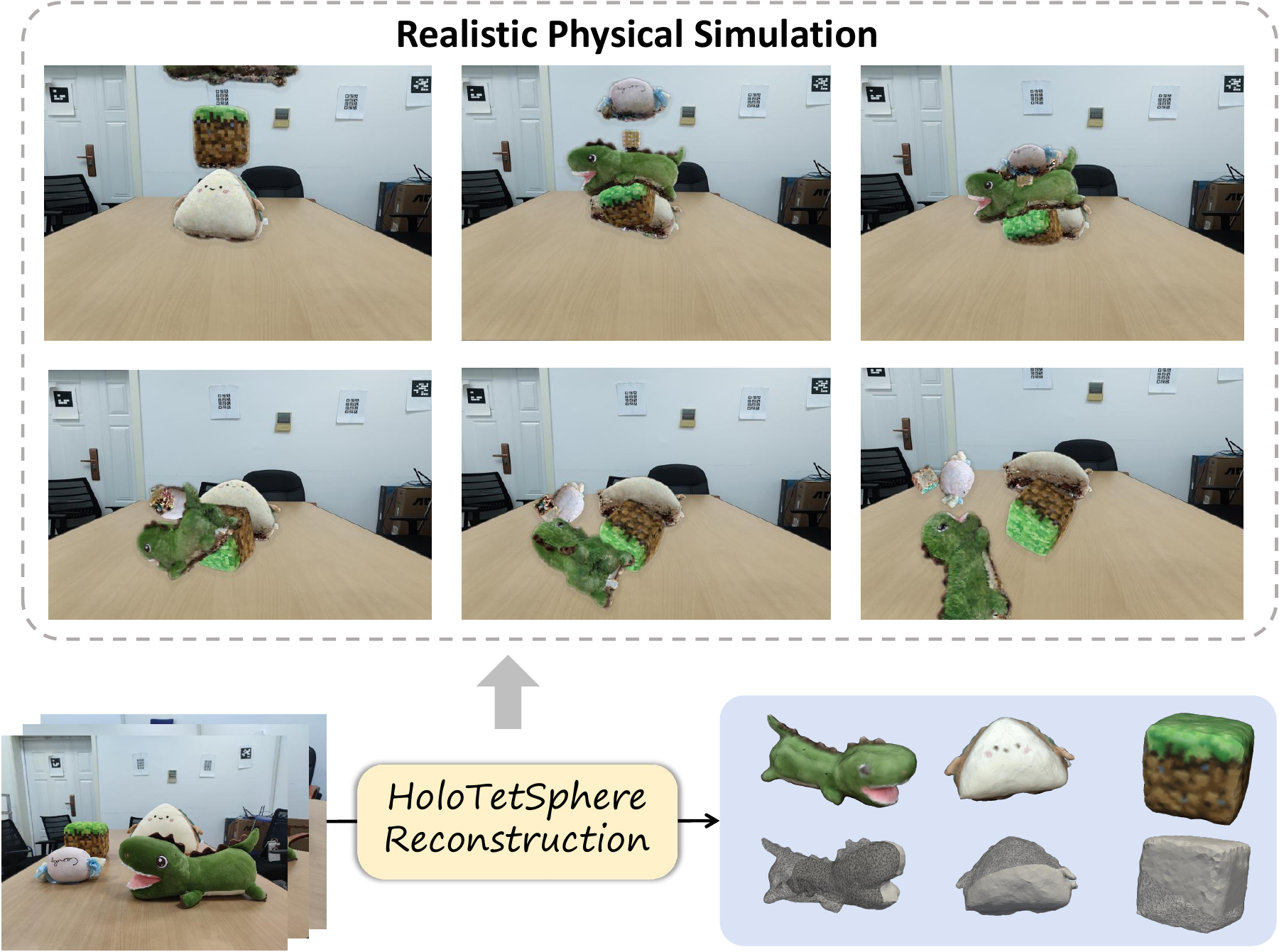}
 \caption{Physical simulations based on our method, which produce accurate geometry and topology of objects from multi-view observation
 in a representation suitable for surface-level physical simulation after the lightweight element cleanup described in the main paper.
 }
  \label{fig:physical_sim}
    \vspace{-6px}
\end{figure}

Additionally, to evaluate the effectiveness of our method at the scene level, we conduct experiments on real-world multi-object datasets. Specifically, we initialize the scene by enclosing multiple objects within a unified convex hull, which allows for the successful reconstruction of the corresponding tetrahedral meshes. Subsequently, the individual object meshes are segmented through clustering to facilitate downstream physical simulations. As illustrated in \cref{fig:physical_sim}, this pipeline enables both high-quality reconstruction and stable physical simulation, from which we further synthesize rendered images based on the simulated tetrahedral mesh coordinates. More results can be found in the supplementary video.
 
% =====================================================================
\section{Limitations and Future Work}
\label{sec:limitations}
% =====================================================================

\subsection{Detailed Limitation Analysis}

\paragraph{Viewpoint completeness.}
Our initialization requires computing a convex hull from the Gaussian
point cloud, which in turn relies on sufficient coverage of the object from
the input views.  Objects photographed from only a narrow angular range
may yield a degenerate convex hull that does not enclose the full geometry,
degrading subsequent tetrahedral optimization.

\paragraph{Open-surface objects.}
The method assumes that reconstructed objects are \emph{solid}, i.e.\ that
a valid volumetric interior exists.  Open thin-shell structures such as
garments (present in the DeepFashion3D dataset) violate this assumption:
the interior of the clothing provides no photometric observations, causing
the opacity field to prune interior tetrahedra incorrectly and preventing
the network from recovering the correct open surface.
For such objects, a watertight proxy is still produced, but it will over-
smooth the fine surface wrinkles that only the outer surface captures.
Future work should explore topology-aware initialization strategies that
begin from an open surface rather than a convex hull for thin-shell inputs.

\paragraph{Thin structures.}
Very thin, plate-like structures are challenging for a volumetric tetrahedral
representation: recovering them requires the tetrahedra to collapse toward a
shell, which conflicts with the orientation-preserving regularization that
resists degenerate/inverted elements. To delineate this reliability boundary,
we reconstruct cubes compressed to decreasing thickness and measure the
recovered thickness between opposing surfaces. As shown in
\cref{tab:thin_supp}, reconstruction stays accurate down to a relative
thickness (compression ratio) of about $0.34$ (relative error $\le 11.8\%$),
whereas below this threshold the recovered thickness deviates sharply
(e.g.\ $85.1\%$ error at ratio $0.17$) as the structure becomes too thin to
support a stable volumetric interior. This provides practical guidance on the
applicable range of our method.

\begin{table}[t]
\centering
\scriptsize
\caption{Thickness stress test. ``Err.'' is the relative thickness error;
``Ratio'' is the relative thickness (compression ratio) with respect to the
uncompressed cube.}
\label{tab:thin_supp}
\begin{tabular*}{\linewidth}{@{\extracolsep{\fill}}cccc}
\toprule
\textbf{GT thickness} & \textbf{Reconstructed thickness} & \textbf{Err.} & \textbf{Ratio} \\
\midrule
0.5126 & 0.5116 & 0.2\%   & 1.00 \\
0.4315 & 0.4297 & 0.4\%   & 0.84 \\
0.3482 & 0.3536 & 1.6\%   & 0.67 \\
0.1761 & 0.1969 & 11.8\%  & 0.34 \\
0.0883 & 0.1634 & 85.1\%  & 0.17 \\
\bottomrule
\end{tabular*}
\end{table}

\paragraph{Scene-level reconstruction.}
The scene-level extension relies on a
clustering step to decompose the scene into individual objects before
per-object tetrahedral reconstruction.  The current clustering relies on
gaussian cues and has not been evaluated quantitatively; its
robustness in cluttered scenes with heavy occlusion remains an open
question.

\subsection{Failure Cases}
Our method may produce degraded results in two scenarios.

\noindent\textbf{Extremely thin structures.} Thin elongated features (e.g.\
antenna-like protrusions below 2–3 voxels wide) may be pruned
prematurely, as the barycentric averaging of the opacity field causes the
entire slender tetrahedron to receive a low opacity even if it corresponds
to genuine surface.

\noindent\textbf{Highly transparent objects.} For glass-like materials,
low photometric opacity is a genuine visual property rather than an
indicator of empty space.  Opacity-based pruning may therefore
inadvertently remove valid tetrahedra, leading to incomplete reconstructions.

\subsection{Future Directions}
Promising directions for follow-on work include:
(i)~\emph{Adaptive tetrahedral refinement} to concentrate mesh resolution
on high-curvature surface regions;
(ii)~\emph{Dynamic scene reconstruction}, extending the pipeline to
non-rigid objects with known or learned deformation fields;
(iii)~\emph{Open-surface-aware initialization} to handle thin-shell
objects without requiring a solid interior assumption; and
(iv)~\emph{Real-time reconstruction} via parallel GPU-accelerated
tetrahedral optimization.

\end{document}